\documentclass[cup6a]{cupconf}
\RequirePackage{graphicx}%
\RequirePackage{epsf}%

\title[DM chemical abundances in nebulae]{Ionized gaseous nebulae chemical abundance determination 
using the direct method}

\author[E. P\'erez-Montero]{Enrique P\'erez-Montero}

\affiliation{Instituto de Astrof\'\i sica de Andaluc\'\i a - CSIC. Apdo. 3004, 18080, Granada, Spain}

\newcommand{\hii}{H\,{\sc ii}\rm}

\newcommand{\siii}{[S\,{\sc iii}]}
\newcommand{\nii}{[N\,{\sc ii}]}
\newcommand{\oiii}{[O\,{\sc iii}]}
\newcommand{\oii}{[O\,{\sc ii}]}
\newcommand{\sii}{[S\,{\sc ii}]}

\newcommand{\neiii}{[Ne\,{\sc iii}]}

\newcommand{\apj}{ApJ}
\newcommand{\apjl}{ApJL}
\newcommand{\apjs}{ApJS}
\newcommand{\aap}{A\&A}
\newcommand{\mnras}{MNRAS}
\newcommand{\aj}{AJ}

\begin{document}
\maketitle

\begin{abstract}
In this tutorial it is explained the procedure to analyze an optical emission-line spectrum
produced by a nebula ionized by massive star formation. Particularly, it is described the
methodology used to derive physical properties, such as electron density and temperature, and
the ionic abundances of the most representative elements whose emission lines are present
in the optical spectrum.
The tutorial is focused on the direct method,based on the measurement
of the electron temperature to derive the abundances, given that the
ionization and thermal equilibrium of the ionized gas 
is dominated by the metallicity. The ionization
correction factors used to obtain total abundances from the abundances
of some of their ions are also given.
Finally, some strong-line methods to derive abundances are described.
These are used when no
estimation of the temperature can be derived,
but that can be consistent with the direct method if they are
empirically calibrated.

\end{abstract}

\section{Introduction}

The determination of chemical abundances is one of the goals of the analysis
of the emission-line spectra produced by the ionization of a gas cloud by massive stars.
By deriving the relative amount of metals in the gas it is possible to give valuable constraints to the past chemical enrichment and
star formation history as these elements can only be produced by nucleosynthesis in the star-cores
before their ejection into the interstellar medium. The relations between metallicity and other observational
properties in the studied objects have important implications in different scales and environments. 
For instance, the integrated metallicity of
star-forming galaxies correlates with stellar mass or star formation rate. Its distribution
across galactic radii in spiral galaxies depends on the evolution of the disks, 
the search for the fraction of primordial helium can constrain other important
cosmological parameters (e.g. \cite{ppl02,peimbert2}),
and in planetary nebulae or envelopes or massive stars the study of chemical abundances allows us to analyze
stellar evolution.

\begin{figure*}
\begin{center}
\includegraphics[width=10cm]{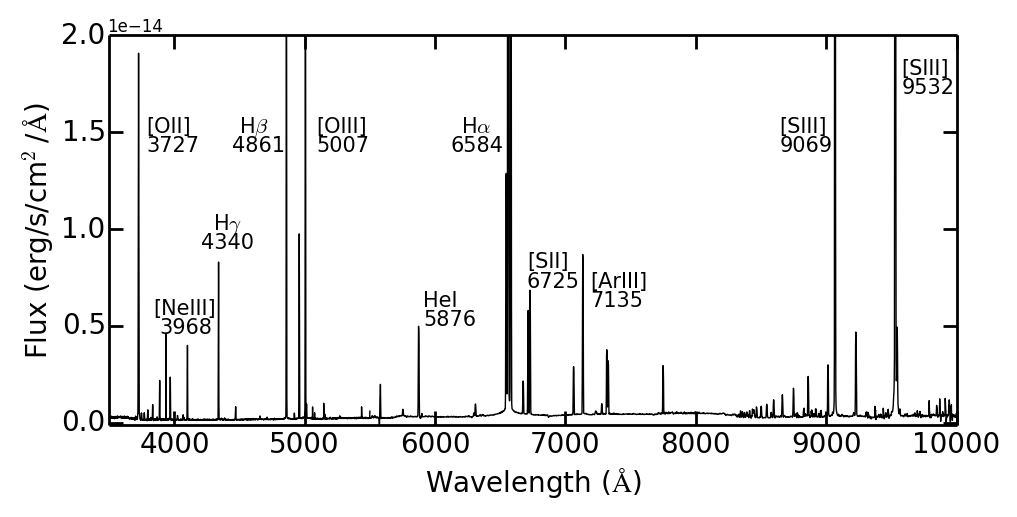}

\end{center}

\caption{Optical spectrum of the Galactic \hii\ region S-156 as taken with the ISIS
spectrograph mounted on the William Herschel Telescope (WHT) (\cite{fdez16})}

\label{spectrum}
\end{figure*}

The processes of massive star formation irradiate light at all
wavelengths,  including  the energetic UV and X-ray able to pull up
the electrons in the atoms of the surrounding gas clouds.
As a consequence the atoms of the gas are excited and ionized,
forming a plasma with free electrons, protons and ions.
When the protons and ions recapture the free electrons of the resulting plasma,
they re-emit the radiation under the form of bright lines
as the electrons fall from level to level towards their ground state 
orbiting around nuclei. These electron recombinations are
in equilibrium with the constant process of photoionization.
Since the typical electron temperature of the nebular plasma is of the same order
than the photons at the optical wavelength
the lines emitted by the gas are prominent and many
times dominate the luminosity of the ionized gas and even of entire
starburst galaxies. 
In Fig. \ref{spectrum} it can be seen a typical emission-line
optical spectrum of a \hii\ region and in Table \ref{lines} a list of the most 
prominent emission lines with their wavelengths can be found.

Among the different emission lines that can be found in an \hii\ region spectrum,
there are the recombination lines (RLs), weakly dependent on the nebular internal temperature. 
Most of the brightest lines emitted by the
lightest elements (H, He) are of this type.
In the case of metals (i.e. all elements heavier than He) RLs provide more
precise abundance determinations (i.e. with a precision better than 5\%)  
but they are are very weak (i.e. around 10$^{-4}$ times
fainter than H$\beta$ 4861 \AA) and thus difficult to measure in weak or
distant objects. Instead, collisional excited lines (CELs) are much brighter and easier
to be detected in optical spectra. These lines correspond to transitions forbidden owing to
heir very low quantum probabilities, but that dominate the cooling of 
the gas under the extremely low densities of the gas in these nebulae.
The intensities of these CELs depend exponentially on the temperature. In principle, the 
temperature can be determined from appropriate line intensity ratios which however 
require  the  detection and measurement of intrinsically faint (or absent)
auroral lines. This is in particular the case for regions with high metal content --
for which the cooling produced by the metals is very efficient and the lines are
no longer detected -- and also for distant \hii\ regions and regions with low surface 
brightness.  
In Table \ref{lines} it is also listed
the nature of the emission lines usually observed in the optical spectrum,
attending to RLs, nebular CELs and auroral CELs.

In those cases where no auroral lines are measured with confidence in the
optical spectrum, other methods based on the strong nebular line
intensities are used. 
However one must take into account
that in many cases, calibrations are obtained from photoionization models
for which the uncertainties are difficult to quantify and lead to absolute metallicity
values not always compatible with the values obtained from the direct method.
See for instance \cite{ke08} to see differences between direct method metallicities and
other obtained from strong-line methods based on models. In contrast, see \cite{dors11} or
\cite{pm14} to see other sets of models that lead to chemical abundances
not systematically higher than those derived from the direct method.

In any case, even when the electronic 
temperature can be determined with high precision, there are some problems
limiting the confidence on the attained results following the
direct method. Those problems include:
(1) the effect of the internal ionization structure on multiple zone models (\cite{pmd03}); 
(2) temperature fluctuations across the nebula (\cite{peimbert03}); 
(3) collisional and density effects on the ionic temperatures  (\cite{lpp99}); 
(4) neutral gas zones affecting the determination of ionization correction factors 
   (ICFs)(\cite{ppl02});
(5) the ionization structure is not adequately described by present models (\cite{pmd03}); 
(6) possible photon escape affecting low ionization lines in the outer regions of
    the nebula (\cite{cdtt01}).The first three effects can introduce uncertainties regarding the derived O abundance
of some 0.2, 0.3 and  0.4 dex respectively, depending on the degree of excitation.
The uncertainties associated to the rest of the enumerated problems have not yet been
quantified.

In this tutorial some very basic instructions to derive physical
properties and ionic chemical abundances following the direct method are
given. The basic assumptions behind these calculations are mainly two:
i) the gas is ionized by hot massive stars whose spectral every distribution can
be modelized by a black-body and ii) all the emission from the ionized gas,
including its complete ionization structure (i.e. from the
innermost layer irradiated by the star up to the photodissociation region, where the gas
is neutral or the outer edge is reached) is well traced by
the observed spectrum.

For this work, all $n_e$, $T_e$ and ionic abundances
were re-calculated using expressions derived
using non-linear fittings to the results obtained from
the emission-line analysis software {\sc pyneb} v0.9.3 (\cite{pyneb}) 
as described below and
with the most updated sets of atomic coefficients
The expressions were obtained using arbitrary sets of 
input emission-line intensities covering the most common conditions
and some of them can be also found in \cite{pm14} or \cite{dors16}.
These formulae are provided to ease the reproducibility
of the calculations, the error analysis and their applicability for large
data samples using different software.

\begin{table}

\label{lines}
\caption{List and properties of the most prominent optical emission lines, including 
wavelength, ion, nature (RL for recombination line, CEL ($n$) for nebular collisional line and
CEL ($a$) for collisional auroral line) and extinction coefficient using the law by \cite{ccm89}.}

\begin{center}
\begin{tabular}{lccc|lccc}
\hline
$\lambda$  (\AA) & Ion & Class & f($\lambda$ & $\lambda$ (\AA) & Ion & Class & f($\lambda$) º\\
\hline

3726.0 & \oii & CEL -($n$)  & 0.322 & 5875.6 & He{\sc i} & RL & -0.203 \\
3728.8 & \oii & CEL - ($n$) & 0.322 & 6300.3 & [O{\sc i}] & CEL ($n$) & -0.263 \\
3868.8 & \neiii & CEL ($n$) & 0.291 & 6312.1 & \siii & CEL ($a$) & -0.264 \\
3967.5  & [Ne{\sc iii} & CEL ($n$) & 0.266 & 6363.8 & [O{\sc i}] & CEL ($n$) & -0.272 \\
3970.0 & H7 & RL & 0.256 & 6548.1 & \nii & CEL ($n$) & -0.295 \\
4068.7 & \sii & CEL ($a$) & 0.239 & 6563.0 & H$\alpha$ & RL & -0.298 \\
4076.4 & \sii & CEL ($a$) & 0.237 & 6583.5  & \nii & CEL ($n$) & -0.304 \\
4102.0 & H$\delta$ & RL & 0.229 & 6678.0 & He{\sc i} & RL & -0.313 \\
4340.0 & H$\delta$ & RL & 0.157 & 6716.4 & \sii & CEL ($n$) & -0.318 \\
4363.2 & \oiii & CEL ($a$)  & -0.149 & 6730.8 & \sii & CEL ($n$) & 0.320 \\
4471.5 & He{\sc i} & RL & 0.10.115 & 7065.0 & He{\sc i} & RL & -0.364 \\
4668.1 & [Fe{\sc iii}] & CEL ($n$) & 0.058 & 7135.8 & [Ar{\sc iii}] & CEL ($n$) & -0.378 \\
4686.0 & He{\sc ii} & RL & 0.050& 7319.5 & \oii & CEL ($a$) & -0.398 \\
4711.3 & [Ar {\sc iii}] & CEL ($n$) & 0.042 & 7330.2 & \oii & CEL ($a$) & -0.400 \\
4713.1 & He{\sc i} & RL & 0.042 & 9017.4 & H{\sc i} P10 & RL & -0.590 \\
4740.1 & [Ar {\sc iv}] & CEL ($n$) & 0.038 & 9068.6 & \siii & CEL ($n$) & -0.594 \\
4861.0 & H$\beta$ & RL & 0.000 & 9231.5 & H{\sc i} - P9 & RL & -0.605 \\
4958.9 & \oiii & CEL ($n$) & -0.026 & 9530.6 & \siii & CEL ($n$) & -0.625 \\
5006.9 & \oiii & CEL ($n$) & -0.038 & 9548.6 & H{\sc i} P8 & RL & -0.626 \\
5754.6 & \nii & CEL ($a$) & -0.185 \\

\hline
\hline

\end{tabular}
\end{center}
\end{table}

\section{Emission-line measurement and extinction}

Once an optical spectrum is conveniently reduced, calibrated and extracted, the analysis
of the emission lines must lead to the measurement of their integrated fluxes
in units of erg$\cdot$s$^{-1}$$\cdot$cm$^{-2}$, reddening corrected and
usually relative to an H{\sc i} line, usually H$\beta$.

The main source of uncertainty associated to the 
derived physical properties are related with the error 
of the fluxes of the lines. The measure of these fluxes 
can be either done manually (e.g. using the {\sc iraf task
{\sc splot}) or using other automatic routines that fit
a Gaussian function to the line profile.} 	
These programs can integer the intensity of each
line over a local continuum. The errors of the 
fluxes measured in this way can be calculated 
with the expression from \cite{glezd94}:

\begin{equation}
\sigma_{l} = \sigma_{c} \cdot \sqrt { N + \frac{EW}{\Delta}} 
\end{equation}
\\

\noindent where
$\sigma_{l}$ is the error of flux of the line, $\sigma_{c}$ represents the
standard deviation in a box near of the measured emission line and 
represents the error in the position of the continuum, $N$ is the 
number of pixels in the measure of the flux of the line, $EW$ is 
the equivalent width of the line and 
$\Delta$ is the dispersion of the wavelength in angstroms per pixel
If a Gaussian function fitting is used to integer the fluxes, this expression 
should be added quadratically to the uncertainty of the fitting, as well as with
other known sources of uncertainty, as flux calibration error.
The use of any automatic routine to measure emission-line fluxes
should consider all these sources of uncertainty in the final error balance.

A previous treatment of the underlying stellar continuum can help
to reduce the uncertainty, above all if the recombination H and He lines are
clearly affected by stellar absorption. Hence, an appropriate subtraction
of this emission before the measurement of the emission lines is convenient
using fitting of synthetic stellar populations (e.g. {\sc starlight}, \cite{starlight}).

\begin{figure*}
\begin{center}

\hspace*{-1.0cm}
\includegraphics[angle=0,width=7cm,clip=]{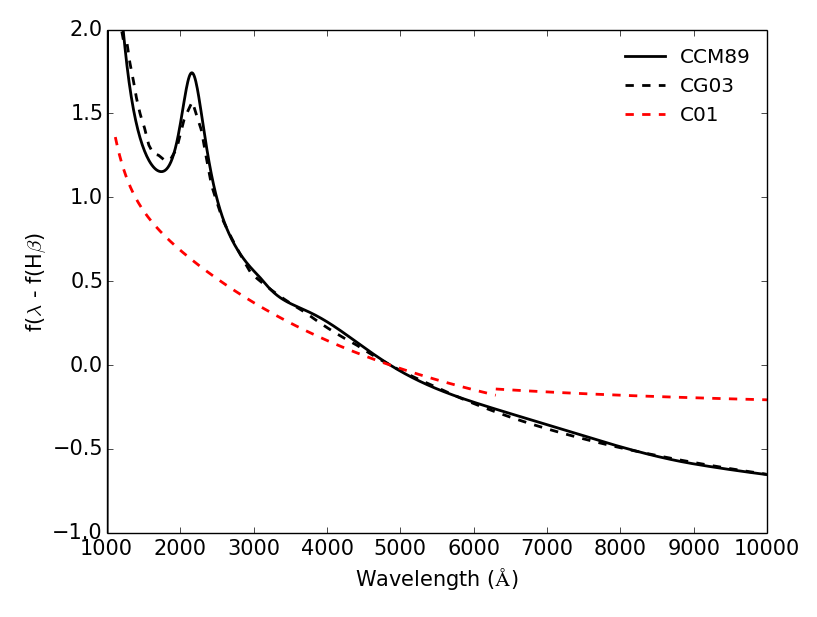}
\includegraphics[angle=0,width=7cm,clip=]{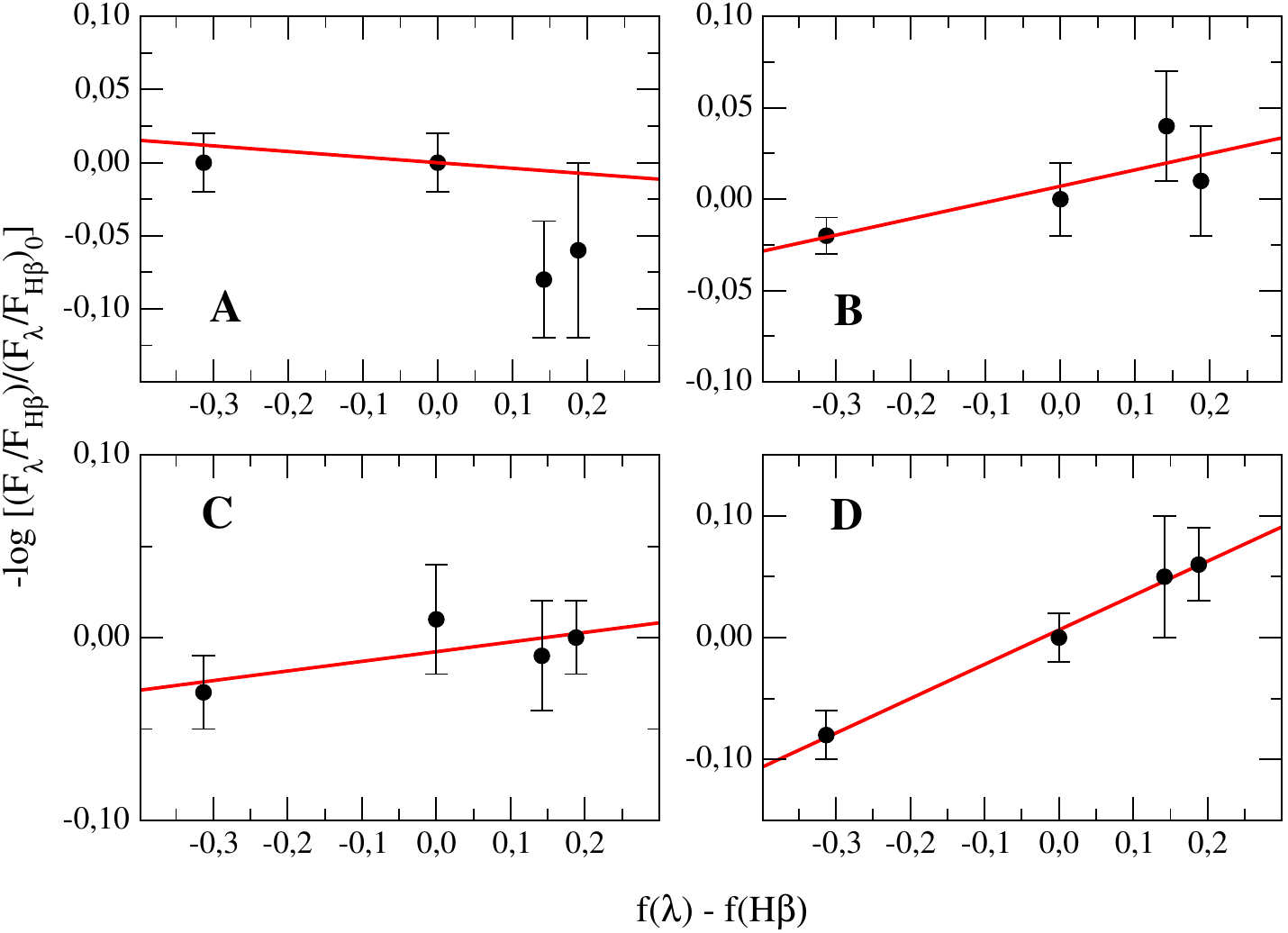}

\end{center}
\caption{At left, different extinction laws including \cite{ccm89}
(solid black line), \cite{gordon03} for the LMC (black dashed line),
and \cite{calzetti} (red dashed line). At right, linear fittings to 
the theoretical-to-observed
H{\sc i} emission lines to obtain the reddening constants in four knots of star
formation in the galaxy IIZw71 (\cite{iizw71}).}∫

\label{extinction}
\end{figure*}

Another important source of uncertainty that can affect the resulting
abundances from the direct method is owing to extinction correction.
The gas is mixed with dust grains that absorb and scatter part of the optical
emission to re-emit it in the infrared affecting the relative
emission line fluxes.
Hence, each one of the intensities measured in this way must be reddening corrected. 
The physical law that describes the extinction of a certain
emission line at a certain $\lambda$ in units of \AA\ is:

\begin{equation}
F_0(\lambda) = 10^{- \tau_\lambda} \cdot F_{obs}(\lambda) =  10^{(1+c(H\beta)) \cdot f(\lambda)}\cdot F_{obs}(\lambda)
\end{equation}
\\

\noindent where $F_0$ is the corrected emission line flux, $F_{jobs}$ is the measured
flux and $\tau_{\lambda}$ is the optical depth at a certain 
wavelength,   $c(H\beta)$ is the reddening constant, what defines the amount of
extinction at the wavelength of H$\beta$. Finally, $f(\lambda)$ is the extinction law, what defines
the behavior of the extinction as a function of wavelength. In left panel of
Fig. \ref{extinction} can be seen
different extinction laws including \cite{ccm89} for the Milky Way,
\cite{gordon03} for the Large Magellanic Cloud, or \cite{calzetti} for starburst galaxies.
These laws present differences above all in the ultraviolet part of the spectrum.
In Table \ref{lines} can also seen the values of $f(\lambda) - f(H\beta)$
according to \cite{ccm89}, assuming a value of $R_V$ = 3.1
(i.e. $R_V$ is defined as the ratio $A_V/(A_V-A_B)$ and $A_V$ and
$A_B$ are the extinctions in mag for bands V and B, respectively).

The reddening constant
$c(H\beta)$ can be calculated from the decrement of Balmer of the stronger recombination lines of hydrogen.
We can write this expression in order to calculate the reddening correction
as a function of the H$\beta$ flux as:

\begin{equation}
\frac{I_0(\lambda)}{I_0(H\beta)} = \frac{I_{obs}(\lambda)}{I_{obs}(H\beta)} \cdot 10^{c(H\beta) \cdot [f(\lambda)-f(H\beta)]}
\end{equation}
\\

Then, $c(H\beta)$ can be derived comparing the observed flux ratios
between the most prominent hydrogen recombination lines and the theoretical
expected values, which depend on density and temperature. In Table \ref{recomb}
are listed the expected values between the flux of H$\beta$ and other H{\sc i} lines
as predicted using data from \cite{storey95} in the case B (i.e. photons are
absorbed as soon as they are emitted).
Using this same data set \cite{lopezs15} provide a polynomial
fitting of the theoretical Balmer ratios as a function of temperature in units of K
for a fixed density of 100 cm$^{-3}$. the resulting fittings are the following:

\begin{equation}
\frac{{\rm I}(H\alpha)}{{\rm I}(H\beta)} = = 10.35 - 3.254 \cdot \log  T_e + 0.3457 \cdot(\log T_e)^2
\end{equation}

\begin{equation}
\frac{{\rm I}(H\gamma)}{{\rm I}(H\beta)} = 0.0254 + 0.1922 \cdot \log  T_e - 0.0204\cdot(\log T_e)^2
\end{equation}

\begin{equation}
\frac{{\rm I}(H\delta)}{{\rm I}(H\beta)} = -0.07132 + 0.1436 \cdot \log T_e - 0.0153\cdot(\log T_e)^2
\end{equation}
\\

\noindent although other ratios for different atomic data can also be
found in \cite{of}.

Then the reddening constant can be derived using
a linear fitting, as in Fig. \ref{extinction} or for doublets of lines
as of H$\alpha$ and H$\beta$ when the other recombination H{\sc i} lines
are very faint.
Since the calculation of reddening depends on temperature and density, if the
derived temperature deviates much from the adopted value, it is better to
make an iterative analysis to check the consistency of the calculated extinction.
When the resulting reddening constant is negative, it is usually
assumed that no dust attenuation is produced and no reddening correction
is applied to the relative emission lines.

The uncertainty associated with the reddening correction must be taken into the account in the errors of the emission lines and
conveniently propagated into all the physical magnitudes we calculate.
Nevertheless, one way to minimize this uncertainty is to correct a given emission line relative
to its closest H{\sc i} line in wavelength and then using the theoretical ratio to H$\beta$.

In the case of integrated spectra from galaxies or extragalactic \hii\ regions, 
one has to take care that the mechanism of ionization is not different
than massive stars (e.g. active galactic nucleus, shocks, post-AGB stars)
with a spectral energy distribution that can make the ionization
equilibrium to be different. 
In these cases the derivation of the extinction, the
physical properties or the chemical abundances using this prescription is not possible.
Normally AGNs can be 
identified by means of X-ray emission or the use of adequate diagnostic diagrams 
(e.g. BPT \cite{bpt}, see also \cite{kewley06}).
 See also \cite{sanchez15} for a description of other
methods of selection of star-forming regions.

\begin{table*}
\label{recomb}
\caption{Theoretical H{\sc i} flux ratios as a function of density and
temperature using \cite{storey95} data under case B approximation.}
\begin{center}
\begin{tabular}{lcccc|cccc}
\hline
\hline
   & \multicolumn{4}{c}{n$_e$ = 100 cm$^{-3}$} & \multicolumn{4}{c}{n$_e$ = 1000 cm$^{-3}$} \\
Ratio & 7\,500 K & 10,000 K & 15,000 K & 20,000 K & 7,500 K & 10,000 K & 15,000 K & 20,000 K \\
\hline
H$\alpha$/H$\beta$ & 2.93 & 2.86 & 2.79 & 2.75 & 2.92 & 2.86 & 2.78 & 2.74 \\
H$\gamma$/H$\beta$ & 0.464 & 0.468 & 0.473 & 0.475 & 0.465 & 0.469 & 0.473 & 0.475 \\
H$\delta$/H$\beta$ & 0.256 & 0.259 & 0.262 & 0.264 & 0.256 & 0.259 & 0.263 & 0.264 \\
P10/H$\beta$ &  0.0189 & 0.0184 & 0.0177 & 0.0172 & 0.0189 & 0.0184 & 0.0177 & 0.0172 \\
P8/H$\beta$ & 0.0376 & 0.0366 & 0.0350 & 0.0339 & 0.0376 & 0.0366 & 0.0350 & 0.0339\\
\hline
\end{tabular}
\end{center}
\end{table*}

\section{Electron temperature and density}

\begin{figure*}
\begin{center}
\hspace*{-1.0cm}
\includegraphics[angle=0,width=7cm,clip=]{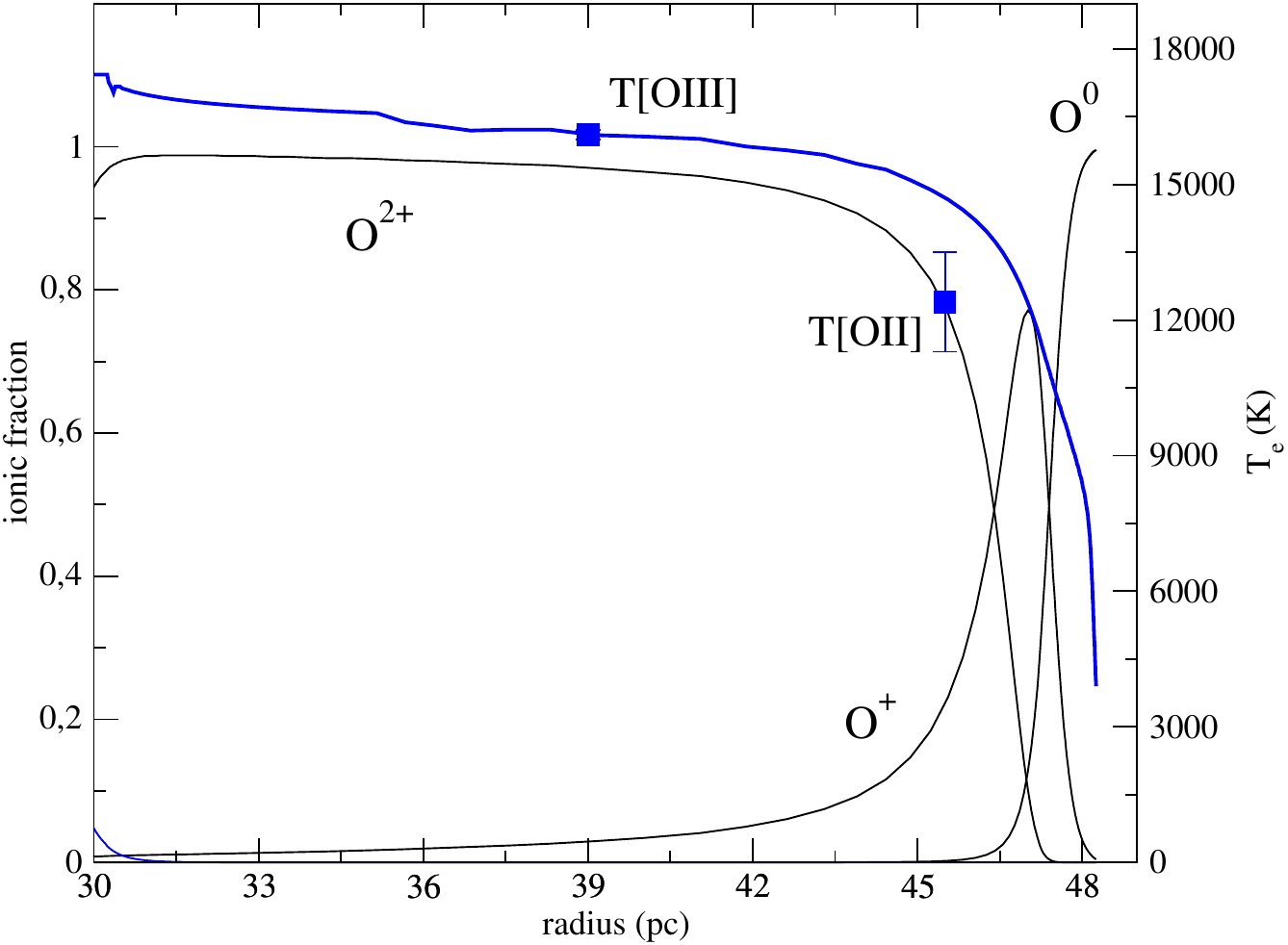}
\includegraphics[angle=0,width=7cm,clip=]{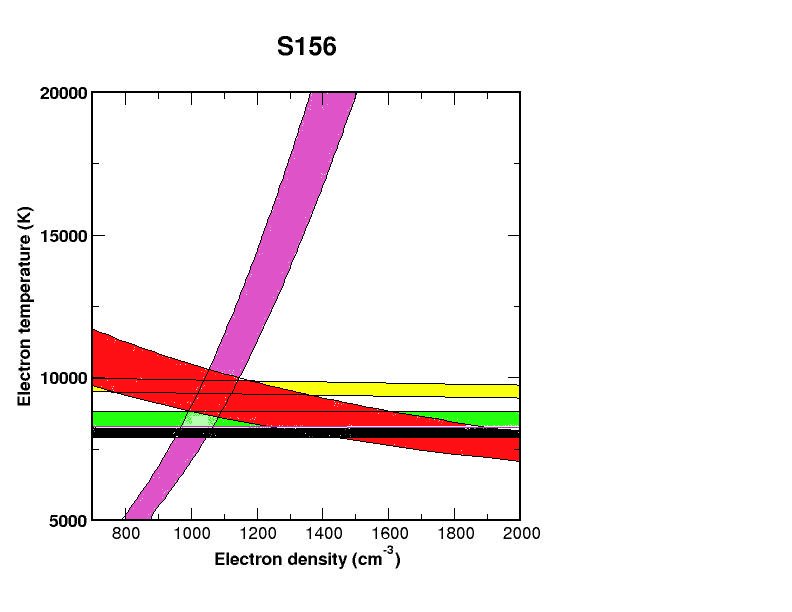}

\caption{Two examples of the thermal and density structure of the gas in \hii\ regions:
Left, radial profile of oxygen abundances and temperature in the brightest knot of the
blue compact dwarf galaxy Mrk~209 as derived from models
(\cite{pmd07}). Right: Diagnostic empirical $n-T$ diagram for the \hii\
region Sh-156 \cite{fdez16}.
Blue band corresponds to n(\sii), black to t(\oiii), yellow to t(\nii), 
red to t(\sii), and green to t(\siii).}

\label{thermal}
\end{center}
\end{figure*}

The thermal and density structure of \hii\ regions can be obtained by
means of the measurement of the fluxes of their emission-lines. 
The direct method relies completely on the measurement of at least one auroral
emission line and the radial thermal structure can be derived using as
many measured temperatures as possible or assuming the temperature of the
different zones from the measured values.
In Fig. \ref{thermal} we see two examples of thermal structure, one owing to
photoionization models and the other to the measurement of different line
diagnostics in a $t-n$ diagram.
The relation
between these temperatures is not many times simple and depends on factors such
as density profile, dust-to-gas ratio, geometry of the gas (matter or density-bounded),
so the most accurate method is to get as many temperature diagnostics as possible to
derive consistently the chemical abundances of each ion.

One simple sketch of the thermal inner structure is the one proposed
by \cite{garnett92} assuming three different zones: i) the {\em high-excitation}
zone, which is the innermost, and corresponds to t(\oiii) ($t_h$), ii) the
{\em low-excitation} zone, which is the outermost, and corresponds to 
electron temperatures such as  t(\oii),
t(\sii) or t(\nii) ($t_l$). Finally, iii) the {\em intermediate excitation} zone appears between
these two and it is mainly represented by t(\siii) ($t_m$). 

Another limitation to calculate the physical properties is the dependence
of the derived electron temperatures on density and vice versa. This is usually overcome using
an iterative method if the assumed initial conditions deviate much from the results, or using
diagnostic diagrams as the one shown in right panel of Fig. \ref{thermal}.

In this section, it is described
the expressions to derive them from the strongest collisional lines usually observed in
the optical spectrum.

\subsection{Oxygen}

\begin{figure*}
\begin{center}
\hspace*{-1.0cm}
\includegraphics[angle=0,width=7cm,clip=]{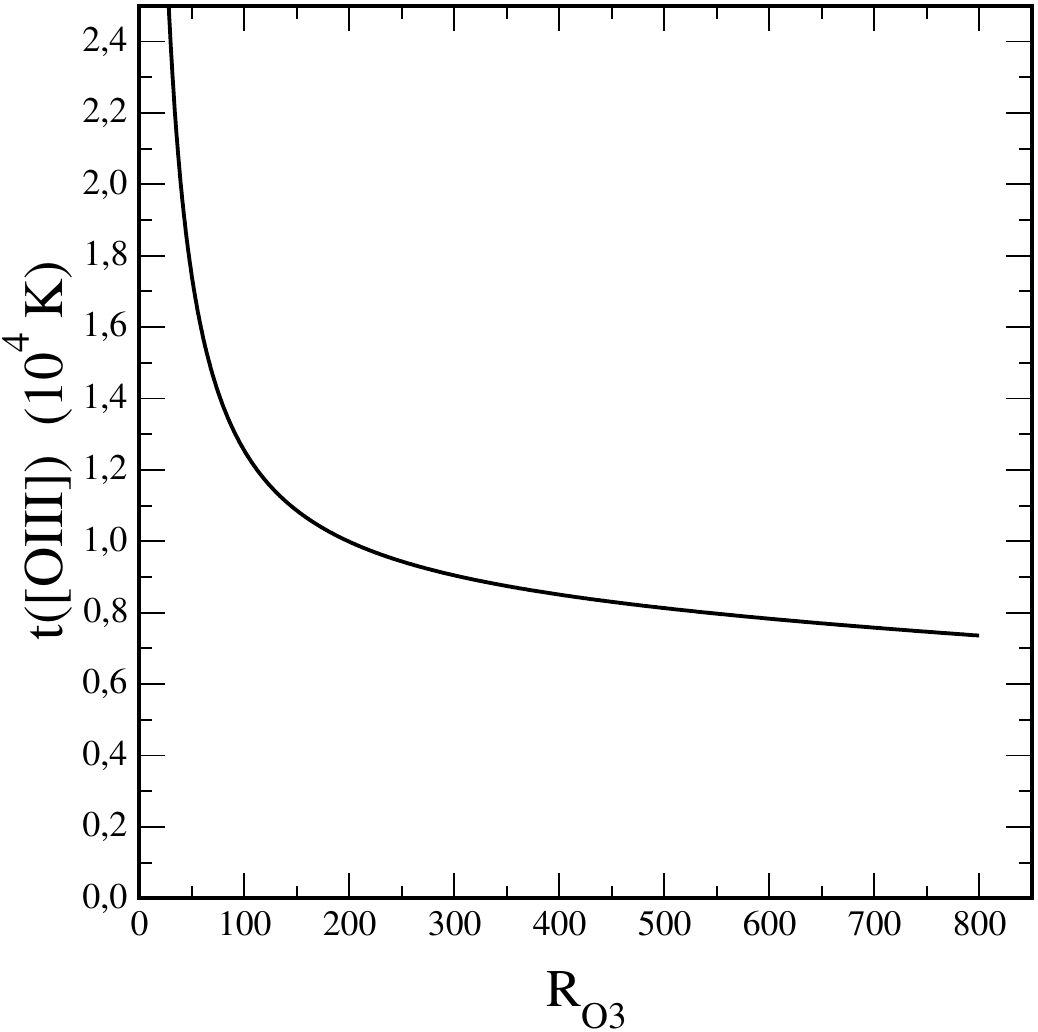}
\includegraphics[angle=0,width=7cm,clip=]{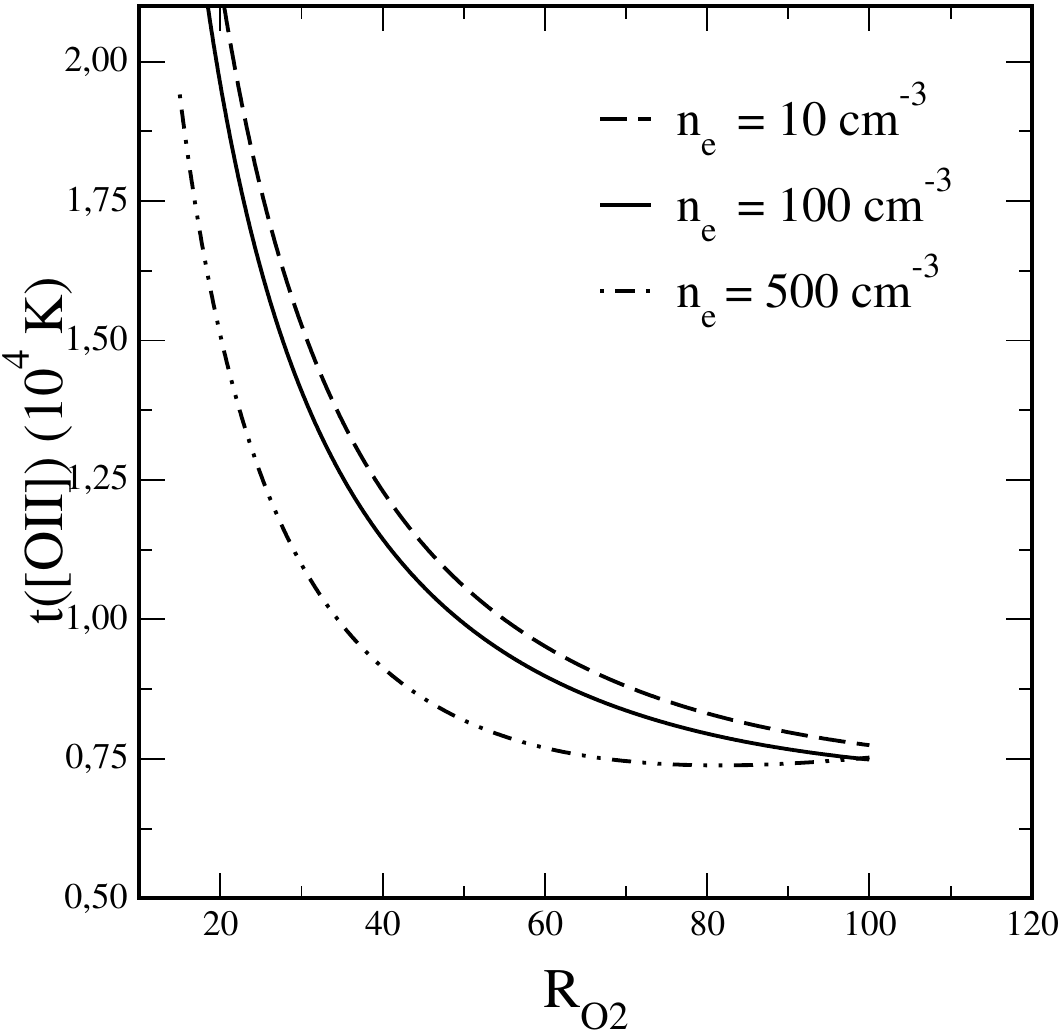}

\end{center}

\caption{Relations between the corresponding nebular-to-auroral lines and the electron
temperature for \oiii\ (left) and \oii (right). In this last case, as a function
of the electron density.} 

\label{RO3}

\end{figure*}

The $T_e$ of \oiii\ can be calculated from the emission-line
ratio: 

\begin{equation}
R_{O3} = \frac{I(4959)+I(5007)}{I(4363)}
\end{equation}
\\

\noindent given that, according to \cite{of}, temperature can be obtained from the 
ratio of collisional  transitions that have a similar energy but occupy different 
levels.  In this calculation it is not required to measure the two strong \oiii\ lines
as there is a theoretical ratio between them (I(5007) = 3$\cdot$ I(4959) ).
The fitting between the ratio and the electronic temperature was obtained
using the program {\sc pyneb} 
assuming a five level atom  
and the following
non-linear fitting for $n_e$ = 100 cm$^{-3}$:

\begin{equation}
t(\textrm{\oiii}) = 0.7840 - 0.0001357\cdot R_{O3} + \frac{48.44}{R_{O3}} 
\end{equation}
\\

\noindent in units of 10$^4$ K
, valid in the range $t$ = 0.7 - 2.5
and using collisional strengths from \cite{ak99}. This fitting gives
precisions better than 1\% for 1.0$<$ t(\oiii) $<$ 2.5, and
better than 3\% for 0.7 $<$ t(\oiii) $<$ 1.0. It was calculated
for a density of 100 cm$^{-3}$, but considering a density
of 1000 cm$^{-3}$ reduces the temperature only in a 0.1\%.
This relation with the resulting fitting can be seen in left panel of Fig. \ref{RO3}.

For [O {\sc ii}] the quotient for the electron temperature is calculated from

\begin{equation}
R_{O2} = \frac{I(3726)+I(3729)}{I(7319\AA)+I(7330\AA)}
\end{equation}
\\

One has to be careful in this case, because the [O {\sc ii}] auroral lines might be
contaminated by recombination emission. Such emission however, can be quantified and 
corrected, according to  \cite{liu2001}, as such contribution can be fitted (for 
 0.5 $\leq$ t $\leq$ 1.0) by the function:

\begin{equation}
\frac{I_R(7319+7330)}{I(H\beta)} = 9.36 \cdot t^{0.44} \cdot \frac{O^{2+}}{H^+}
\end{equation}
\\

Moreover, the ratio of the [O {\sc ii}] lines is strongly dependent on the electron 
density. Ideally, one should know the [O {\sc ii}] density from the ratio 
I(3726\AA)/I(3729\AA) but very frequently we lack resolution to separate the doublet
in which case one has to resort  to the [S{\sc iii}] density, also representing the low
excitation zone. The fitting obtained is:

\begin{equation}
t(\textrm{\oii})=a_0(n)+a_1(n) \cdot R_{O2}+\frac{a_2(n)}{R_{O2}}
\end{equation}
\\

\noindent that also gives $t$ in units of 10$^4$ K and 
where the coefficients are respectively:\\

$$a_0(n) = 0.2526 - 0.000357 \cdot n - \frac{0.43}{n}$$
$$a_1(n) = 0.00136 + 5.42 \cdot 10^{-6} \cdot n + \frac{0.00481}{n}$$
\begin{equation}
a_2(n) = 35.624 - 0.0172 \cdot n + \frac{25.12}{n}
\end{equation}
\\

\noindent being $n$ the electron density in units of cm$^{-3}$ and
using the collisional coefficients from \cite{pradham06} and \cite{tayal07}.
The fittings were made in the range t = 0.8 - 2.5 in units of 10$^4$ K with an
uncertainty better than 2\%. The resulting fittings for
different electron densities can be seen in right panel of Fig. \ref{RO3}.

\begin{figure*}
\begin{center}
\hspace*{-1.0cm}
\includegraphics[angle=-90,width=7cm,clip=]{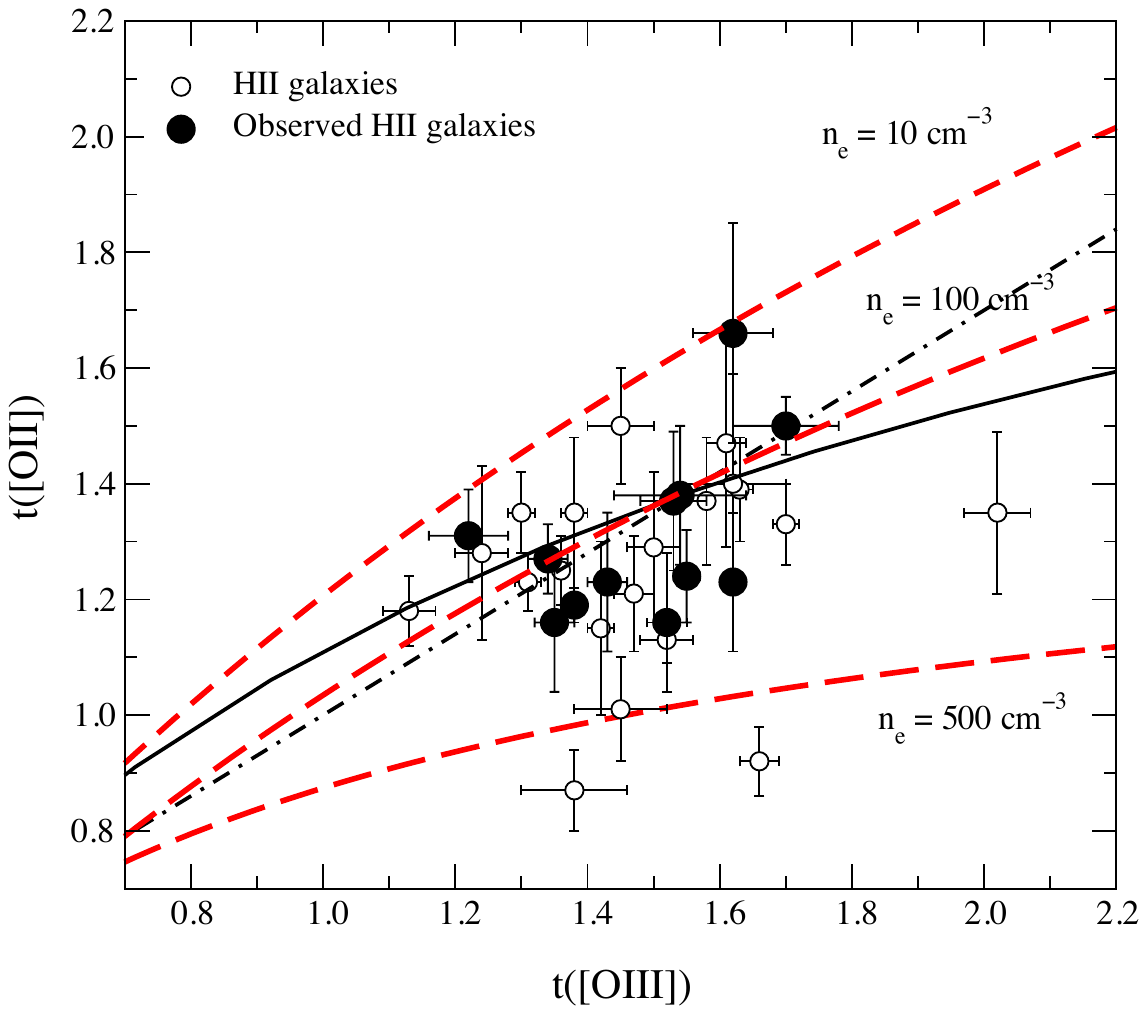}
\includegraphics[angle=-90,width=7cm,clip=]{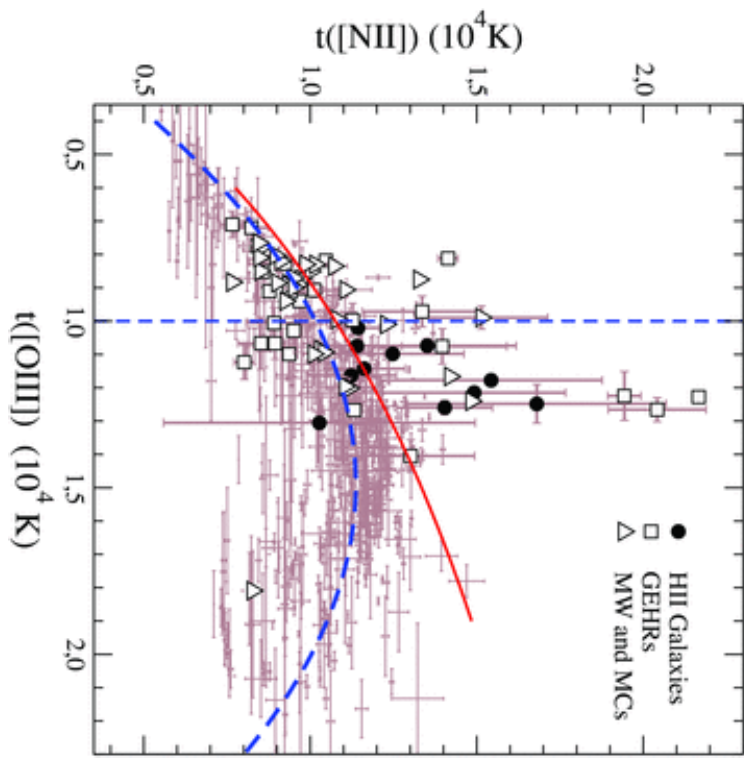}

\hspace*{-1.0cm}
\includegraphics[angle=-90,width=7cm,clip=]{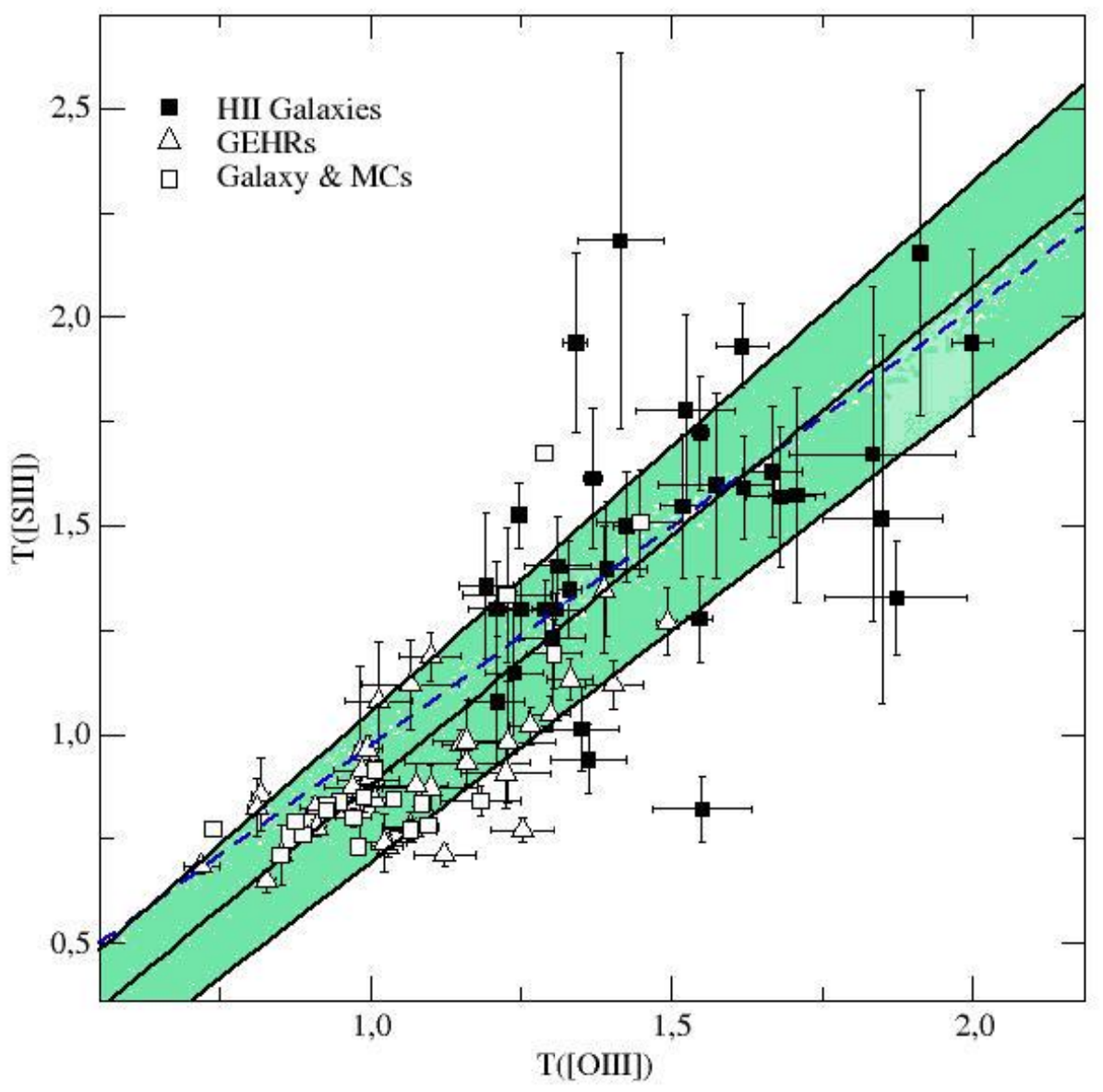}
\includegraphics[angle=-90,width=7cm,clip=]{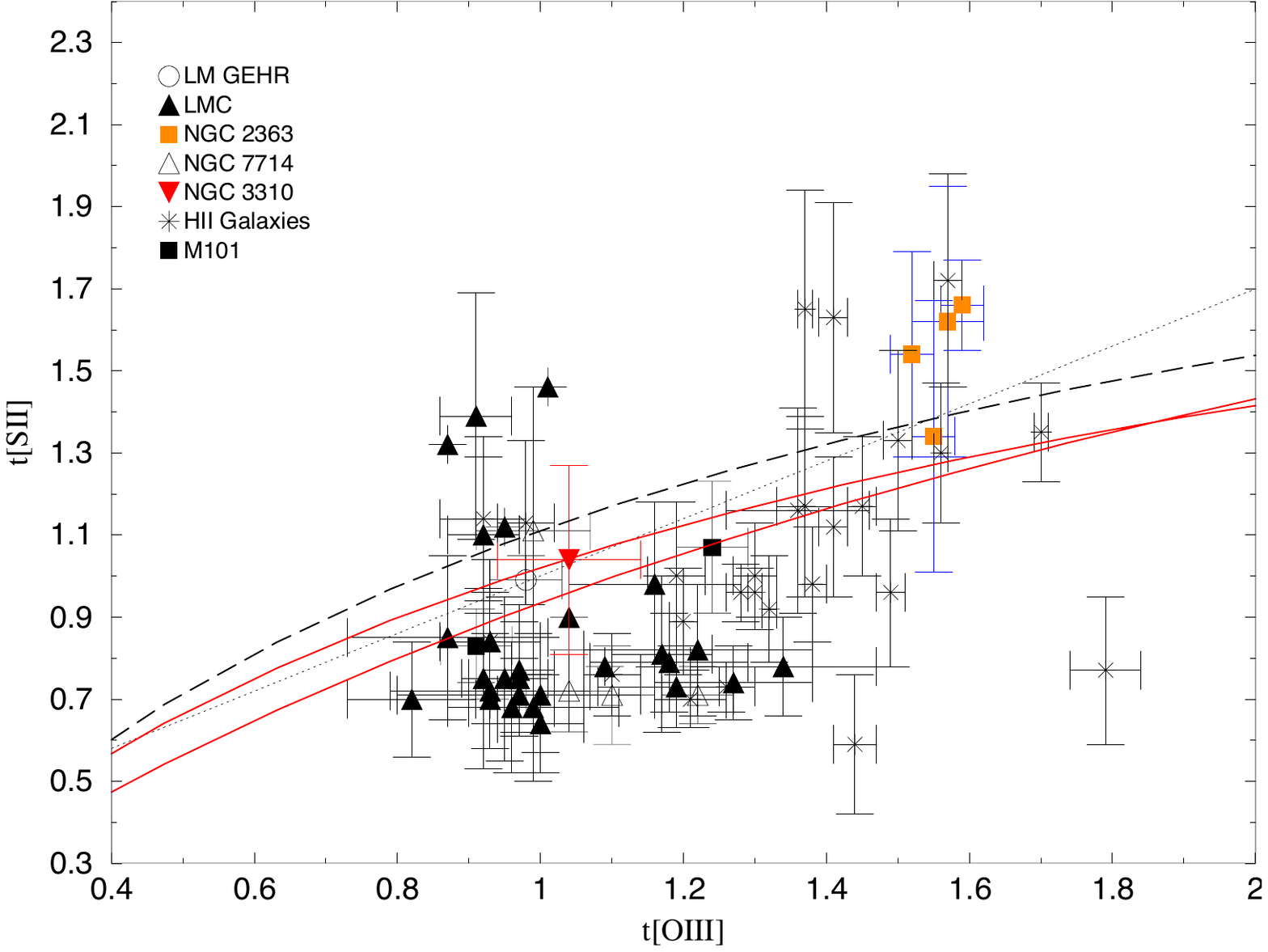}
\end{center}

\caption{Empirical and model-based relations between different line temperatures: 
In the upper left panel, relation between
t(\oiii) and t(\oii) (\cite{pmd03}), in the upper right, relation between t(\oiii) and t(\nii)
(\cite{pmc09}). In the upper right panel, relation between t(\oiii) and t(\siii) (\cite{H06}) and
in the lower right panel, relation between t(\oiii) and t(\sii) (\cite{thesis}).
}

\label{rels}

\end{figure*}

Often the auroral [O {\sc ii}] lines are not observed with good S/N or they are 
outside our observed spectral range. In that case it is practical to use some relation 
based on photoionization models in order to infer t[O {\sc ii}] from t[O {\sc iii}]. For instance,
the relation 

\begin{equation}
t(\textrm{\oii}) = \frac{2}{t(\textrm{\oiii})^{-1}+0.8}
\end{equation}
\\

\noindent based on  \cite{stasinska90} models is frequently accepted.
However, such expression neglects the dependence of t[O {\sc ii}] on the density, 
consistent with the dispersion for the objects for which both temperatures 
have been derived from observations as can be seen in left upper panel of Fig. \ref{rels}. 
In the same panel are shown fittings to grids of models, presented in \cite{pmd03},
with different electron densities.
The density-dependent calibration obtained in this case and given in \cite{H06} is:

\begin{equation}
t(\textrm{\oii}) = \frac{1.2+0.002 \cdot n + \frac{4.2}{n}}{t(\textrm{\oiii})^{-1} + 0.08 + 0.003 \cdot n + \frac{2.5}{n}}
\end{equation}
\\

\subsection{Sulfur}

The [S{\sc ii}] line ratio is commonly used  to determine electron density for 
the low excitation zone.
It is generally assumed that the nebula has constant density, although there is
growing evidence for the existence of a density profile instead.
Fortunately, the high excitation species have diagnostic ratios that are almost no
density sensitive.

Electron densities are necessary for the derivation of chemical abundances
of ions of the type np$^2$, such as O$^+$. These densities were derived using
the following emission-line ratio:

\begin{equation}
R_{S2} = \frac{I(6716)}{I(6731)}
\end{equation}
\\

The following expression
is proposed to derive the electron density:

\begin{equation}
\textrm{n$_e$(\sii)} = 10^3 \cdot \frac{R_{S2} \cdot a_0(t) + a_1(t)}{R_{S2} \cdot b_0(t) + b_1(t)}
\end{equation}
\\

\noindent with $n_e$ in units of cm$^{-3}$ and $t$ in units of
10$^4$ K. 
Using the appropriate fittings and {\sc pyneb} with 
collision strengths from \cite{TZ10} gives these polynomial fittings 
to the coefficients

\[ a_0(t) = 16.054 - 7.79/t - 11.32\cdot t \]
\[ a_1(t) = -22.66 + 11.08/t + 16.02\cdot t \]
\[ b_0(t) = -21.61 + 11.89/t + 14.59\cdot t \]
\begin{equation}
b_1(t) = 9.17 - 5.09/t - 6.18\cdot t 
\end{equation}
\\

This expression fits the density calculated by {\sc pyneb} better
than a 1\% for temperatures in
the range 0.6 $< t_e <$  2.2 and densities in the range 10 $< n_e <$ 1000.

\noindent here $t$ is generally t[O {\sc iii}], although an iterative process could be used to 
calculate it with  t[S{\sc iii}] given that this temperature, like t[O {\sc ii}], a type np$^3$ ion,
is density dependent.
The ratio used in this case is:

\begin{equation}
R_{S2}' = \frac{I(6717)+I(6731)}{I(4068)+I(4076)}
\end{equation}
\\

For the [S{\sc ii}] auroral lines it is enough to measure one of them, as they are related by a fixed theoretical ratio, I(4068\AA) $\approx$ 3$\cdot$ I(4076\AA).
We can calculate in this way the [S{\sc ii}] temperature.

\begin{equation}
t(\textrm{\sii}) = a_0(n) + a_1(n) \cdot R_{S2}' + \frac{a_2(n)}{R_{S2}'} +  \frac{a_3(n)}{R_{S2}'^2}
\end{equation}
\\

\noindent where

$$a_0(n) = 0.99 + \frac{34.79}{n} + \frac{321.82}{n^2}$$
$$a_1(n) = -0.0087 +  \frac{0.628}{n} + \frac{5.744}{n^2}$$
$$a_2(n) = -7.123 + \frac{926.5}{n} - \frac{94.78}{n^2}$$
\begin{equation}
a_3(n) = 102.82 + \frac{768.852}{n} - \frac{5113}{n^2}
\end{equation}
\\

\noindent also fitted with the coefficients by \cite{TZ10} in the range
$t$ = 0.8 to 2.5 in units of 10$^4$ K and $n_e$ up to 500 cm$^{-3}$ with
a precision better than 2\%.

When the [S{\sc ii}] auroral lines are not available, it is usually assumed that
t([S{\sc ii}]) $\approx$ t[O {\sc ii}]. There is evidence however, suggesting a somewhat lower value.
From the models used in \cite{thesis}, it is obtained a lineal fitting:

\begin{equation}
t(\textrm{\sii}) = 0.71 \cdot t(\textrm{\oii}) + 0.12
\end{equation}
\\

\noindent for a 100 cm$^{-3}$ number density. For lower densities, this expression seems to 
be valid. In any case, for the few objects for which we have a simultaneous 
measurement of both temperatures, the dispersion is quite large as can be seen in 
lower left panel of Fig. \ref{rels} for \hii\ galaxies.

Direct measurements of the [S {\sc iii}] temperature become possible with the 
availability of the collisional lines in the near IR.

\begin{equation}
R_{S3} = \frac{I(9069)+I(9532)}{I(6312)}
\end{equation}
\\

This expression can be simplified in case of lacking one of the near 
IR lines, knowing that
I(9532\AA) $\approx$ 2.44 $\cdot$ I(9069\AA).
With this ratio it is possible the corresponding temperature
with the following fitting in the range
$t$ = 0.6 - 2.5 using the collision strengths from \cite{HRS12}

\begin{equation}
t(\textrm{\siii}) = 0.5147 + 0.0003187\cdot R_{S3} + \frac{23.64041}{R_{S3}} 
\end{equation}
\\

\noindent with a precision better than 1\% in the range
0.6 $<$ t(\siii) $<$ 1.5, and better than 3\% up to values
t(\siii) = 2.5. These values enhance in less than a 3\% when the
considered density goes from 100 to 1000 cm$^{-3}$.

As discussed in \cite{garnett92},  t[S {\sc iii}] is in between the temperatures of [O {\sc iii}]
and of [O {\sc ii}] and allows us to calculate the  S$^{2+}$ abundance from just the 6312 \AA\ 
line in high-metallicity objects. 
In case that the [\siii] cannot be measured, it is possible to derive 
t([\siii]) from t(\oiii). anv viceversa
The empirical fitting given by \cite{H06} is:

\begin{equation}
t(\textrm{\siii}) = (1.19\pm0.08) \cdot t(\textrm{\oiii}) - (0.32\pm0.10)
\end{equation}
\\

\noindent which can be seen in left lower panel of Fig. \ref{rels}.

\subsection{Nitrogen}

The$T_e$ of \nii\ can be calculated using the
ratio:

\begin{equation}
R_{N2} = \frac{I(6548)+I(6583)}{I(5755)}
\end{equation}
\\

\noindent that, with the corresponding fitting leads to
the expression:

\begin{equation}
t(\textrm{\nii}) = 0.6153 - 0.0001529\cdot R_{N2} + \frac{35.3641}{R_{N2}} 
\end{equation}
\\

\noindent also in units of 10$^4$ K, in the range $t$ = 0.6 - 2.2
using collision strengths from \cite{T11}. This fit gives a precision better
than 1\% in the range 0.7 $<$ t([\nii]) $<$ 2.2 and better than 3\% in
the range 0.6 $<$ t([\nii]) $<$ 0.7. It was calculated for a density
of 100 cm$^{-3}$, but for a density of 1000 cm$^{-3}$, the temperature
is reduced in less than a 1\%.

The nebular lines of [N{\sc ii}] are very close to H$\alpha$ so they appear sometimes blended to this
line and therefore it is not possible to measure both of them.
In this case it is often 
assumed the following theoretical relation between them, I(6583) $\approx$ 2.9 $\cdot$ I(6548).
Besides, the auroral line of [N{\sc ii}] is affected by recombination emission, that can be corrected using the next expression
proposed by \cite{liu2001}:

\begin{equation}
\frac{I_R(5755)}{I(H\beta)} = 3.19 \cdot t^{0.30} \cdot \frac{N^{2+}}{H^+}
\end{equation}
\\

\noindent in the range between 5\,000 and 20\,000 K.

Unfortunately, the auroral line has very low signal-to-noise ratio, so it is usually considered the 
approximation t([N{\sc ii}]) $\approx$ t([O{\sc ii}]) as valid. This relation is confirmed by photoionization models
but is quite sensitive to density and the inner ionization structure of the nebula, and it is possible to reach values 
closer to t([S{\sc iii}]) in some cases, so in all case it can't be taken without any sort of uncertainty. 

Other possibility is to calculate t([N{\sc ii}]) directly from t([O{\sc iii}]) using
the expression derived using photoionization models by \cite{pmc09},
as can be seen in lower right panel of Fig. \ref{rels}:

\begin{equation}
t(\textrm{\nii}) = \frac{1.85}{t(\textrm{\oiii})^{-1}+0.72}
\end{equation}
\\

\subsection{Balmer temperature}
\begin{figure*}
\begin{center}

\includegraphics[width=8cm,angle=-90]{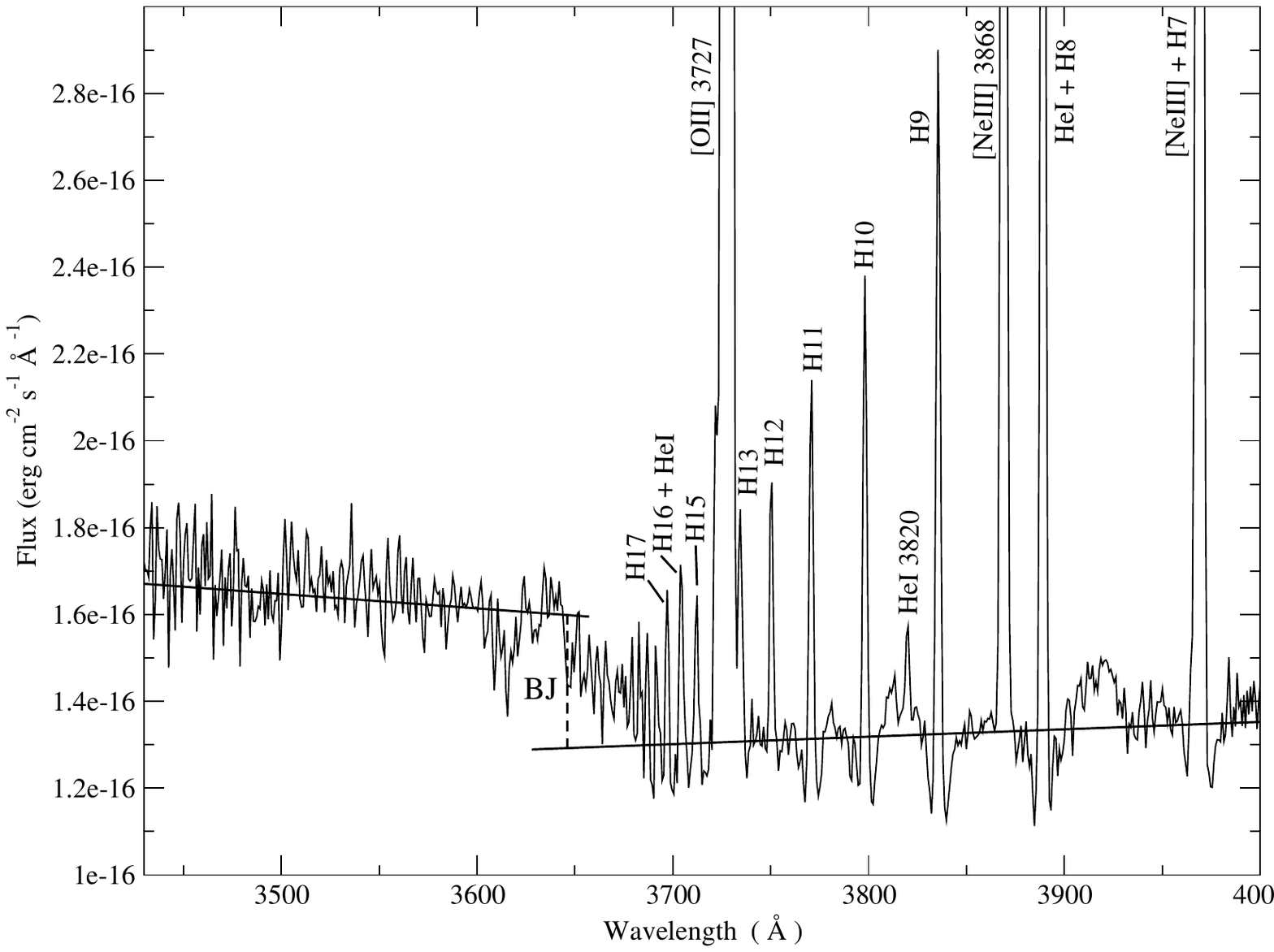}

\caption{Measurement of the Balmer jump in the \hii\ galaxy 
SDSS J003218.60+150014.2 in \cite{H06}.
  The solid lines traces the continuum to both sides of the Balmer jump and the
  dashed line depicts the value of its measurement.} 
\label{Bjump}

\end{center}
\end{figure*}

The Balmer temperature is an alternative to other
temperatures calculated using collisionally excited lines.
It depends on the value of the Balmer 
jump (BJ) in emission (i.e. the continuum nebular emission
at a bluer wavelength than the Balmer series).
In order to measure this value, it 
is necessary to fit the continuum in both sides of the 
discontinuity  ($\lambda_B$\,=\,3646\,\AA). 
An example of the measurement of the BJ in an the optical
spectrum of an \hii\ galaxy can be seen in Fig. \ref{Bjump}.

The contribution 
of the underlying stellar population affects, between other factors, 
to the emission of hydrogen lines near from the BJ. The increment 
in the number of lines at shorter wavelengths produces blends 
between them that trend to reduce the level of the continuum at 
a redder wavelength of the discontinuity so it is necessary to 
take into the account all this in the final uncertainty. Once 
measured the BJ, the Balmer temperature  (T(Bac)) is measured 
from the quotient of the flux of the jump and the emission of 
the line H11 by means of the expression given in \cite{liu2001}:

\begin{equation}
T(Bac)\,=\,368\times(1\,+\,0.259y^+\,+\,3.409y^{++})\Big(\frac{BJ}{H11}\Big)^{-3/2}\,K
\end{equation}
\\

\noindent
where $y^+$ and $y^{++}$ are the ionic abundances of helium 
once and twice ionized, respectively, and BJ is in 
ergs\,cm$^{-2}$\,s$^{-1}$\,\AA$^{-1}$.

\subsection{Other temperatures and densities}

There are other collisional excited lines for other ions in the optical spectrum,
but their corresponding temperatures or densities cannot be directly measured using the appropriate
emission line ratios in the optical or their auroral lines are too weak
to be detected.
This is the case for [Ar{\sc iv}] for instance, for which we can derive the density in the 
inner nebula, or the temperatures of the rest of ions for which the corresponding auroral line are not available in the
observed spectral range or there is not enough signal-to-noise ratio to detect their recombination lines.
In this case, it is usually taken some assumptions about the gas structure 
and the position of each species is associated to another ion whose
temperature is known. This is the case of the next ions, assuming the three-zone
temperature sketch proposed by \cite{garnett92}:

\begin{equation}
t(\textrm{[Ne{\sc iii}]}) \approx t(\textrm{[Fe{\sc iii}]}) \approx t(\textrm{He{\sc ii}}) \approx t(\textrm{[Ar{\sc iv}}]) \approx  t(\textrm{\oiii}) \approx t_h
\end{equation}
\\

\begin{equation}
t(\textrm{[Ar{\sc iii}]}) \approx t(\textrm{\siii}) \approx t_m
\end{equation}
\\

\begin{equation}
t(\textrm{\sii}) \leq t(\textrm{\nii}) \approx t(\textrm{\oii}) \approx t_l
\end{equation}
\\

\section {Ionic and total chemical abundances}

In what follows I will describe the expressions used to derive the chemical abundances of
the ions whose emission lines can be measured in the optical spectrum.
These formulae use the appropriate electron densities and temperatures obtained in the determination
of the thermal structure of the gas.
Most of the fittings follow equations with the same mathematical relations
than those proposed by \cite{pagel92}.

In addition, in order to derive total chemical abundances,
the ionization correction factor for each species will be also described.
ICFs are used to calculate
total abundances from the observed ions of that element in the
optical range, that usually do not fulfill all the ionic stages present in the gas.

\begin{equation}
\frac{N(X)}{N(H)} = ICF(X_{obs}) \cdot \frac{N(X_{obs})}{N(H^+)} 
\end{equation}
\\

\subsection{Helium}

Helium lines, as equal as hydrogen ones in the 
visible spectrum, have a recombination nature and 
they are generally strong and numerous, although 
many of them are usually blended with other lines. 
Besides they are affected by absorption of underlying 
stellar population, by fluorescence or by collisional 
contribution. 

The total He abundance can be calculated in the optical spectrum using
lines of He{\sc i} and He{\sc ii}.
Following the same equations as described in \cite{os04}, 
Once subtracted the stellar absorption and reddening correction has been carried out,
the abundance of He$^+$ can be derived for each He{\sc i} RL as:

\begin{equation}
y^+(\lambda) = \frac{\rm{I}(\lambda)}{\rm{I}(H\beta)} \frac{F_{\lambda}(n,t)}{f_{\lambda}(n,t,\tau)}
\end{equation}
\\

\noindent where $F_{\lambda}(n,t)$ 
is the theoretical emissivity scaled to H$\beta$ and 
$f_{\lambda}$ isthe optical depth function. 
It is generally used the strongest lines, including
He{\sc i}\,$\lambda\lambda$\,4471, 5876, 6678 and 
7065\AA\ and then calculating $y^+$ using a weighted mean,
but a precise determination of He abundance requires the use of as
many lines as possible (see \cite{ppl02}).

Using {\sc pyneb} with the atomic data from \cite{porter13a} 
the theoretical
emissivities related to H$\beta$ can be calculated:

\[
F_{4471} = (2.0301 + 1.5 \cdot 10^{-5} \cdot n) \cdot t^{0.1463 - 0.0005 \cdot n}
\]
\[
F_{5876} = (0.745 - 5.1\cdot 10^{-5}\cdot n)\cdot t^{0.226 - 0.0011\cdot n}
\]
\[
F_{6678} = (2.612 - 0.000146 \cdot n) \cdot t^{0.2355 - 0.0016\cdot n}
\]
\begin{equation}
F_{7065} = (4.329 - 0.0024 \cdot n) \cdot t^{-0.368 - 0.0017 \cdot n}
\end{equation}
\\

\noindent with $t$ in units of 10$^4$ K and $n$ in units of cm$^{-3}$
and calculated for temperatures between 8\,000 and 25\,000 K
and densities between 10 and 1000 cm$^{-3}$. The precision of the fittings is
better than 1\% in all case.
Normally it is taken the temperature of \oiii, as it is the most precise.
The optical depth functions can be taken from \cite{os04},
but these are only important for high-extinction objects and when it is required
a very precise measurement of this abundance (e.g. for the search for
the primordial helium abundance).

The abundance of He$^{2+}$ can be calculated using 
the He{\sc ii} 4686 \AA\ line and a high-excitation
temperature by means of the recombination coefficients by \cite{storey95} using the 
following expression:

\begin{equation}
y^{2+} = \frac{\rm{I}(4686)}{\rm{I}(H\beta)} \cdot 0.0416 \cdot t^{-0.146}
\end{equation}
\\

\noindent calculated in the same range as He{\sc i} emissivities.

Then, the total He abundance can be derived
the addition of the abundances of the two ions:

\begin{equation}
y = y^+ + y^{2+}
\end{equation}
\\

\noindent where $y$ is the relative abundance of helium (He/H),
$y^+$ the abundance of He$^+$, and $y^{2+}$ the abundance of He$^{2+}$.

\subsection{Oxygen}
The chemical abundance of O$^+$ can be derived
with the relative intensity of \oii\ 3726, 3729 \AA\
emission lines to H$\beta$ and the corresponding temperature using
the following expression obtained from fittings to {\sc pyneb} using
the default collision strengths from \cite{pradham06} and \cite{tayal07}:

\[
12+\log\left(\frac{O^+}{H^+}\right) = \log\left(\frac{I(3726)+I(3729)}{I(H\beta)}\right) +
+ 5.887 + \frac{1.641}{t_l} - 0.543\cdot\log(t_l) + 
\]
\begin{equation}
+ 0.000114\cdot n_e
\end{equation}
\\

\noindent with a precision better than 0.01dex in the temperature
range 0.7 $<$ t(O$^+$) $<$ 2.5 and density of 100 cm$^{-3}$. For
a density of 1000 cm$^{-3}$ the precision is better than 0.02dex.

Alternatively \cite{kniazev03} suggest the use of the 
7319,7330 {\AA} doublet in those objects observed with a setup
that does not cover the 3727 \oii line.

\[
 12+\log \left(\frac{{\rm O^{+}}}{{\rm H^{+}}}\right) = \log  \left[ \frac{I(7320+7330)}{I{\rm (H\beta)}}\right]+7.21 
 +\frac{2.511}{t_l}-0.422 \cdot \log t_{l} + 
\]
\begin{equation}
+ 10^{-3.40} n_{\rm e} ( 1-10^{-3.44}\: \cdot n_{\rm e})
 \end{equation}  
\\

Regarding O$^{2+}$, its chemical abundance was derived 
using the relative intensity of \oiii\ 4959, 5007 \AA\
emission lines to H$\beta$ and the corresponding temperature using
the following expression obtained from fittings to {\sc pyneb}:

\begin{equation}
12+\log\left(\frac{O^{2+}}{H^+}\right) = \log\left(\frac{I(4959)+I(5007)}{I(H\beta)}\right) +
6.1868 + \frac{1.2491}{t_h} - 0.5816\cdot\log(t_h)
\end{equation}
\\

\noindent with a precision better than 0.01dex in the temperature range
0.7 $<$ t(O$^{2+}$) $<$ 2.5. A change in the density from 10 to 1000 cm$^{-3}$
implies a decrease of less than 0.01dex in the derived abundance.

The total oxygen abundance can be approximated by

\begin{equation}
\frac{O}{H} = \frac{O^++O^{2+}}{H^+}
\end{equation}
\\

\noindent given that due to the charge exchange reaction the relative fractions of neutral 
oxygen and hydrogen are similar:

\begin{equation}
\frac{O^0}{O} = \frac{H^0}{H}
\end{equation}
\\

However, in some high-excitation spectra where the He{\sc ii} 4586 \AA\ is seen
it can be considered that part of the O is under the form of O$^{3+}$. In that case it can be 
assumed that:

\begin{equation}
\textrm{ICF}(O^++O^{2+}) = 1 + \frac{y^{2+}}{y^+}
\end{equation}
\\

\subsection{Sulfur}

The abundances are obtained from the 6717, 6731 {\AA}\ lines for  S$^+$ and by the
9069, 9532 {\AA}\ lines for S$^{2+}$, though for the latter, also the 6312 {\AA}\ line 
can be used using the following expressions:

\begin{equation}
 12+\log \left(\frac{S^{+}}{H^{+}}\right) = \log  \left( \frac{I(6717)+I(6731)}{I(H\beta)}\right)
+5.463 +\frac{0.941}{t_l}-0.37\cdot\log t_l)
\end{equation}
\\

\noindent and

\begin{equation}
12+\log \left(\frac{S^{2+}}{H^{+}}\right) = \log \left[ \frac{I(9069)+I(9532)}{I(H\beta)}\right]
+ 5.983 + \frac{0.661}{t_m}- 0.527\log (t_m).    
\end{equation}
\\

In case that the near-IR \siii lines cannot be measured but the auroral line
at 6312 \AA\ is available with good signal-to-noise, it is possible to derive S$^{2+}$
abundances from the expression:

\begin{equation}
12+\log\left(\frac{S^{2+}}{H^+}\right) = \log\left(\frac{I(6312)}{I(H\beta)}\right) +
6.695 + \frac{1.664}{t_m} - 0.513\cdot\log(t_m)
\end{equation}
\\

These were derived using the same collisional coefficients as used for the derivation
of expressions for temperature. The fittings have a precision better than 0.01 dex in the 
temperature range 0.7 $<$ t 

The ICF for sulfur takes into account the S$^{3+}$ abundance which cannot be 
determined in the optical range. A good approximation is given by \cite{stasinska78}:

\begin{equation}
\textrm{ICF}(S^++S^{2+}) = \left[ 1-\left(\frac{O^{2+}}{O^{+}+O^{2+}}
  \right)^\alpha\right]^{-1/\alpha} 
\end{equation}
\\

\noindent Although it is customary to write this expression as a function of the 
O$^{+}$/(O$^{+}$+O$^{2+}$) ionic fraction, we have reformulated it in terms of 
O$^{2+}$/(O$^{+}$+O$^{2+}$)  since the errors associated to O$^{2+}$ are
considerably smaller than for O$^{+}$. 
and for a sample of objects with observed  [S{\sc iv}] line at 10.5 $\mu$m, it was derived
$\alpha \approx$ 3.27 (\cite{dors16}).

\subsection{Nitrogen}

We  can calculate N$^+$ abundance from the 6548 and 6583 {\AA}\ lines.
The expression to calculate $N^+/H^+$ abundance using these lines and
assuming a low-excitation temperature is:

\begin{equation}
12+\log\left(\frac{N^+}{H^+}\right) = \log\left(\frac{I(6548)+I(6583)}{I(H\beta)}\right) +
6.291 + \frac{0.90221}{t_l} - 0.5511\cdot\log(t_l) 
\end{equation}
\\

\noindent with a precision better than 0.01dex in the temperature
range 0.6 $<$ t $<$ 2.2. It decreases less than 0.01dex when
the considered density goes from 100 to 1000 cm$^{-3}$.

One can calculate quite precisely the total abundance of Nitrogen
assuming that:

\begin{equation}
\frac{N^+}{N} = \frac{O^+}{O}
\end{equation}
\\

\noindent owing to the similarity of ionization potentials of O$^+$ and $N^+$
,what leads to the corresponding ICF:

\begin{equation}
\textrm{ICF}(N^+) = \frac{O}{O^+}
\end{equation}
\\	

The ratio N$^+$/O$^+$ can be derived directly from the expression:

\begin{equation}
\log\left(\frac{N^+}{O^+}\right) = \log\left(\frac{I(6583)}{I(3726)+I(3729)}\right) +
0.493 - 0.025\cdot t_l - \frac{0.687}{t_l} + 0.1621\cdot\log(t_l) 
\end{equation}
\\

\subsection{Neon}
Ne has [{Ne{\sc iii}] prominent emission lines in the blue part of the spectrum and can be
a good tracer of metallicity as it is not depleted into dust grains.
Since one of the emission lines (3968 \AA) usually appears blended with
and H{\sc i} recombination line (H7), the Ne$^{2+}$ ionic
abundance can be derived using the following expression from the line at 3869 \AA\ and the electron
temperature of the high-excitation zone (usually t(\oiii)):

\begin{equation}
12+\log\left(\frac{Ne^{2+}}{H^+}\right) = \log\left(\frac{I(3869)}{I(H\beta)}\right) +
6.947 + \frac{1.614}{t_h} - 0.4291\cdot\log(t_h) 
\end{equation}
\\

\noindent with a precision better than 2\% in the range of temperature from 0.6 to 2.2 in
units of 10$^4$ K. The collisional coefficients are from \cite{MB00}.

The ionization correction factor for neon can be calculated according to 
the expression given by \cite{pmh07} based on photoionization models: 

\begin{equation}
\textrm{ICF}(Ne^{2+})\,= 0.753 + \,0.142\,x+ \frac{0.171}{x}
\end{equation}
\\

\noindent where $x$\,=\,O$^{2+}$/(O$^{+}$+O$^{2+}$). 
This expression deviates from the classical approximation 
Ne/O$\approx$Ne$^{2+}$/O$^{2+}$ used to derive total neon abundances.

\subsection{Argon}

For argon, we use the [Ar{\sc iii}] 7137 {\AA}\ line. It is possible to measure as well the lines of [Ar{\sc iv}]
at 4713 and 4740 \AA. Nevertheless, the first of them usually appears blended with another line of He{\sc ii} at 4711\AA\ that is
difficult to correct, so it is better to use the second and brighter to calculate the ionic abundance of Ar$^{3+}$.

\begin{equation}
12+\log\left(\frac{Ar^{2+}}{H^+}\right) = \log\left(\frac{I(7135)}{I(H\beta)}\right) +
6.100 + \frac{0.86}{t_m} - 0.404\cdot\log(t_m) 
\end{equation}
\\

\begin{equation}
12+\log\left(\frac{Ar^{3+}}{H^+}\right) = \log\left(\frac{I(4740)}{I(H\beta)}\right) +
6.306 + \frac{1.232}{t_h} - 0.703\cdot\log(t_h) 
\end{equation}
\\

\noindent both with a precision better than 2\% in the fitting in the same temperature range
than in Ne using the collisional coefficients from \cite{GMZ95} for Ar$^{2+}$ and \cite{RB97}
for Ar$^{+3}$.

As in the case of neon, the total abundance of argon can be calculated using
the ionization correction factors (ICF(Ar$^{2+}$) and the
ICF(Ar$^{2+}$+Ar$^{3+}$)) given by \cite{pmh07}. It can be used
the first one only when we cannot derive a value for Ar$^{3+}$. The
expressions for these ICFs are:   

\begin{equation}
\textrm{ICF}(Ar^{2+})\,= 0.596 +\,0.967\,(1-x)+\frac{0.077}{(1-x)}
\end{equation}
\\

\begin{equation}
\textrm{ICF}(Ar^{2+}+Ar^{3+})\,= 0.929+ + \,0.364\,(1-x) + \frac{0.006}{(1-x)}
\end{equation}
\\

\noindent where $x$\,=\,O$^{2+}$/(O$^{+}$+O$^{2+}$). 

\subsection{Iron}

The abundance of Fe is usually derived from the absorption stellar features, but
there are some high-excitation [Fe{\sc iii}] lines that can be used to derive
Fe$^{2+}$ abundances in the ionized gas-phase. For instance, from 4658 \AA\ from {\sc pyneb}
it is obtained:

\begin{equation}
12+\log\left(\frac{Fe^{2+}}{H^+}\right) = \log\left(\frac{I(4658)}{I(H\beta)}\right) +
6.288 + \frac{1.408}{t_h} - 0.203\cdot\log(t_h) 
\end{equation}
\\

\noindent using the collisional coefficients of Fe$^{2+}$ from \cite{Fe3} in the range
t = 0.8 - 2.5. The ICF to derive the total Fe abundance was proposed by \cite{icf_fe}:

\begin{equation}
\textrm{ICF}(Fe^{2+}) = \left(\frac{O^+}{O^{2+}}\right)^{0.09}\left[1+\frac{O^{2+}}{O^{+}}\right]
\end{equation}
\\

\section{Strong-line methods calibrated using the DM}

In absence on any of the auroral lines it is not possible to obtain the electron
temperature, and hence to derive the thermal structure and the ionic and total abundances of the elements
whose lines can be measured in the optical spectrum. However, some of the other nebular CELs
can be measured and be used as indicators of the chemical content of the gas.

As quoted above, many of the calibrations of these so-called strong-line methods
have been made in the literature attending to 
different criteria or calibration sample so it is difficult to assess if they
are compatible between them. The most important rule to follow is try
to be consistent if we want to make a comparative analysis. In this section I describe
different strong-line methods calibrated with large samples of objects with a previous determination
of the electron temperature so they are in principle compatible with the direct method.
For the sake of consistency, it is only here cited those calibrations calculated using
the same methodology as described in the above sections and using a very similar sample of
objects, selected only with criterion of having at least one measured electron temperature, be
ionized by massive star formation, and with no restriction on the resulting chemical abundances.
However, it is worth to mention the existence of other calibrations based on the direct method 
such as \cite{pil10}, \cite{pil16} or \cite{marino13}, among others.}

\begin{figure*}
\begin{center}
\hspace*{-1.0cm}
\includegraphics[width=7cm]{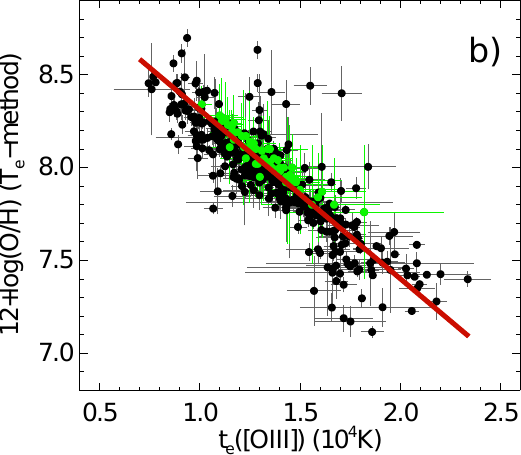}
\includegraphics[width=7cm]{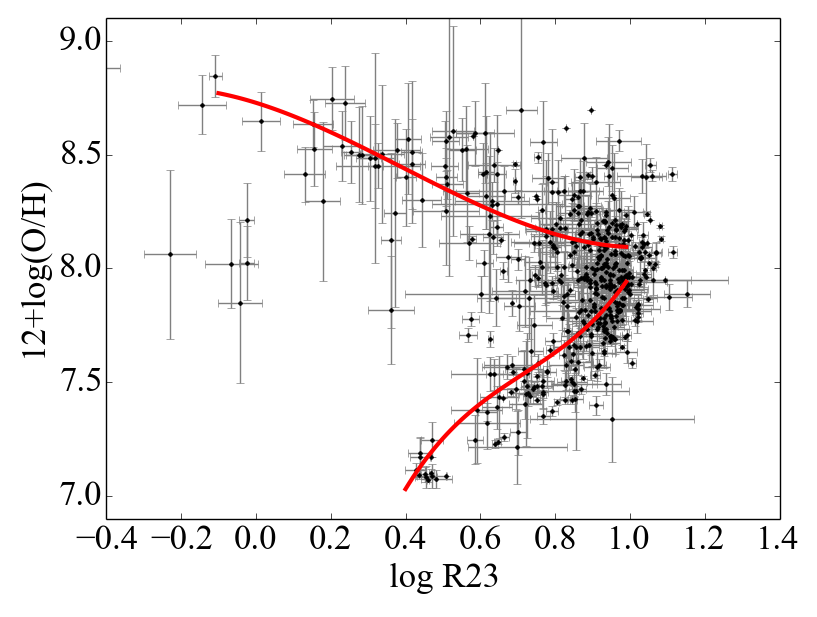}

\caption{Left panel: Empirical relation between electron temperature and
total oxygen abundance (\cite{amorin15}). Right panel, relation between oxygen
abundance and the R23 parameter fro the sample of objects in \cite{pmc09} and
non-linear fittings for the lower and upper branches assuming the average excitation in each range.
}

\label{strong_oh}
\end{center}
\end{figure*}

\subsection{Calibrations of the electron temperature}

Sometimes it is possible to derive an electron temperature and its
corresponding ionic
abundance. Nevertheless, the total abundance cannot be
calculated because the lines corresponding to
other excitation lines cannot be measured. This is the case
of O/H when the \oii\ 3727 \AA\ lines are out of the spectral
range or their corresponding auroral lines at 7319 and 7330 \AA\ are
too faint.

In this case the total oxygen abundance can be estimated directly
from the \oiii\ electron temperature as proposed by \cite{amorin15} and that
can be seen in left panel of Fig.\ref{strong_oh}:

\begin{equation}
12+\log(O/H) = 9.22 (\pm 0.03) - 0.89(\pm 0.02)\cdot t(\textrm{\oiii})
\end{equation}
\\

\noindent with a dispersion of 0.15 dex. This expression can also be used to derive
the temperature from a strong-line derivation of O/H. The electron
temperature is useful for example to quantify the extinction from the Balmer decrement.

\subsection{Parameters based on \oii\ and \oiii\ lines}

The most widely used parameter based on oxygen lines that can be
related to metallicity is R23 given by \cite{pagel79}:

\begin{equation}
R23 = \frac{I(3727)+I(4959,5007)}{I(H\beta)}
\end{equation}
\\

However, there are some known limitations to give a calibration of this
parameter with O/H. The main limitation is that the relation is
double-valued: R23 grows with O/H for low-Z and it decreases for
high-Z, as the main source of cooling when the temperature is lower are
hydrogen recombination lines. In the so-called turnover region
(around 12+log(O/H) $\approx$ 8.0 - 8.3, as can be seen in right panel of
Fig. \ref{strong_oh}) the dispersion is very high and it is very
difficult to provide an accurate determination. 

For this tutorial and for the sake of consistency I provide
fittings to the sample of objects with a direct determination of O/H
in \cite{pmc09}, following the same functional expressions given by
\cite{kob99} based on R23 and on the excitation as a function of
the ratio of \oii\ and \oiii. The fittings for both branches at
an average excitation in each branch are shown in Fig. \ref{strong_oh}.

In the upper branch (12+log(O/H) $>$ 8.25):

\[
12+log(O/H) = 8.656 - 0.411 \cdot x - 0.586 \cdot x^2 + 0.469 \cdot x^3 -
\]
\begin{equation}
- y \cdot (-0.224 + 0.309 \cdot x + 0.741 \cdot x^2 + 0.722 \cdot x^3)
\end{equation}
\\

\noindent where $x$ = log(R23) and $y$ = log(\oii 3727/\oiii 4959+5007).
The dispersion of this fitting, taken as the standard deviation of
the residuals to the abundances from the direct method is 0.17 dex.

For the lower branch (12+log(O/H) $<$ 8.0):

\[
12+log(O/H) = 7.247 - 1.196 \cdot x + 4.078 \cdot x^2 - 2.194 \cdot x^3 - 
\]
\begin{equation}
- y \cdot (1.076 - 3.735 \cdot x + 5.671 \cdot x^2 - 2.947 \cdot x^3)
\end{equation}
\\

\noindent with a dispersion of 0.14 dex.

In cases where the \oiii\ are not observed owing the studied spectral range, 
\cite{pmh07} propose to use the empirical relation with [Ne{\sc iii}] at
3968 \AA\ (= \oiii\ 5007 / 15.37) and using also a hydrogen
recombination line at a blue wavelength, such as H$\gamma$ or H$\delta$.

\begin{figure*}
\begin{center}
\hspace*{-1.0cm}
\includegraphics[width=7cm]{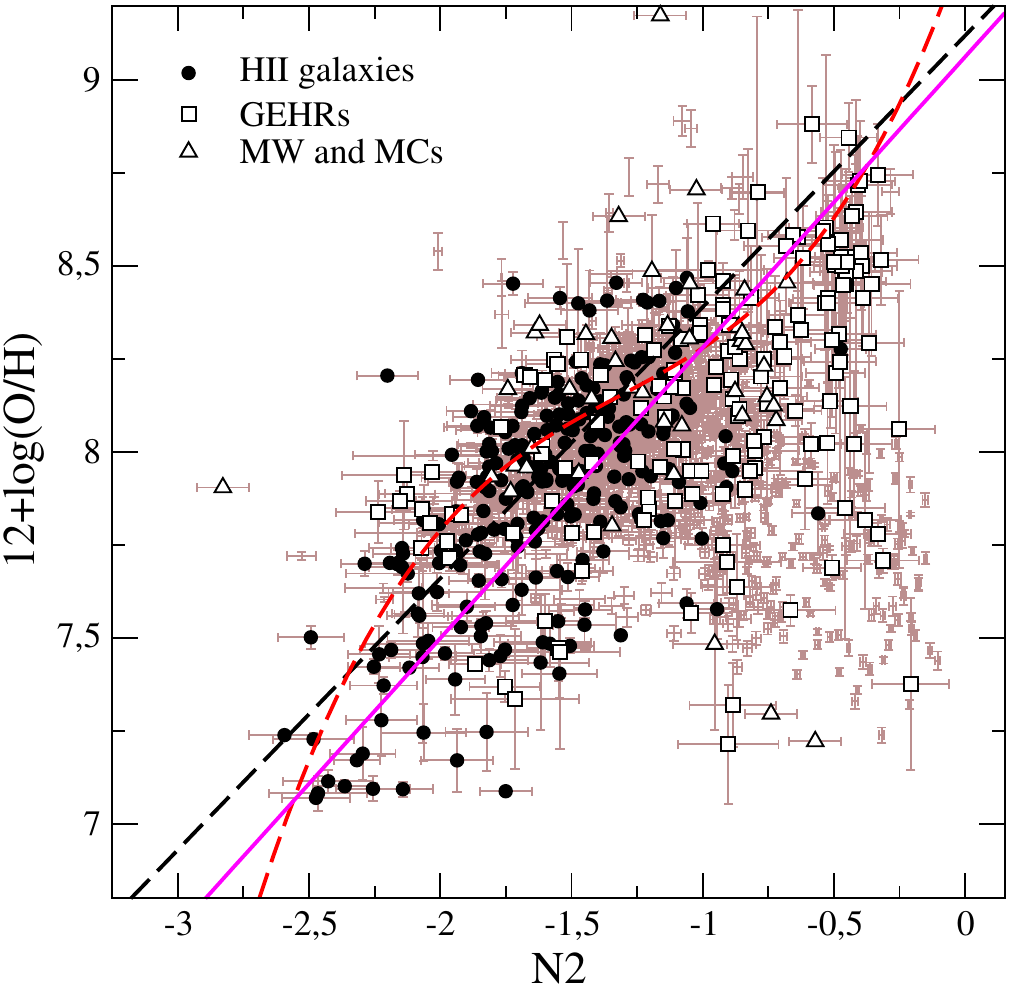}
\includegraphics[width=7cm]{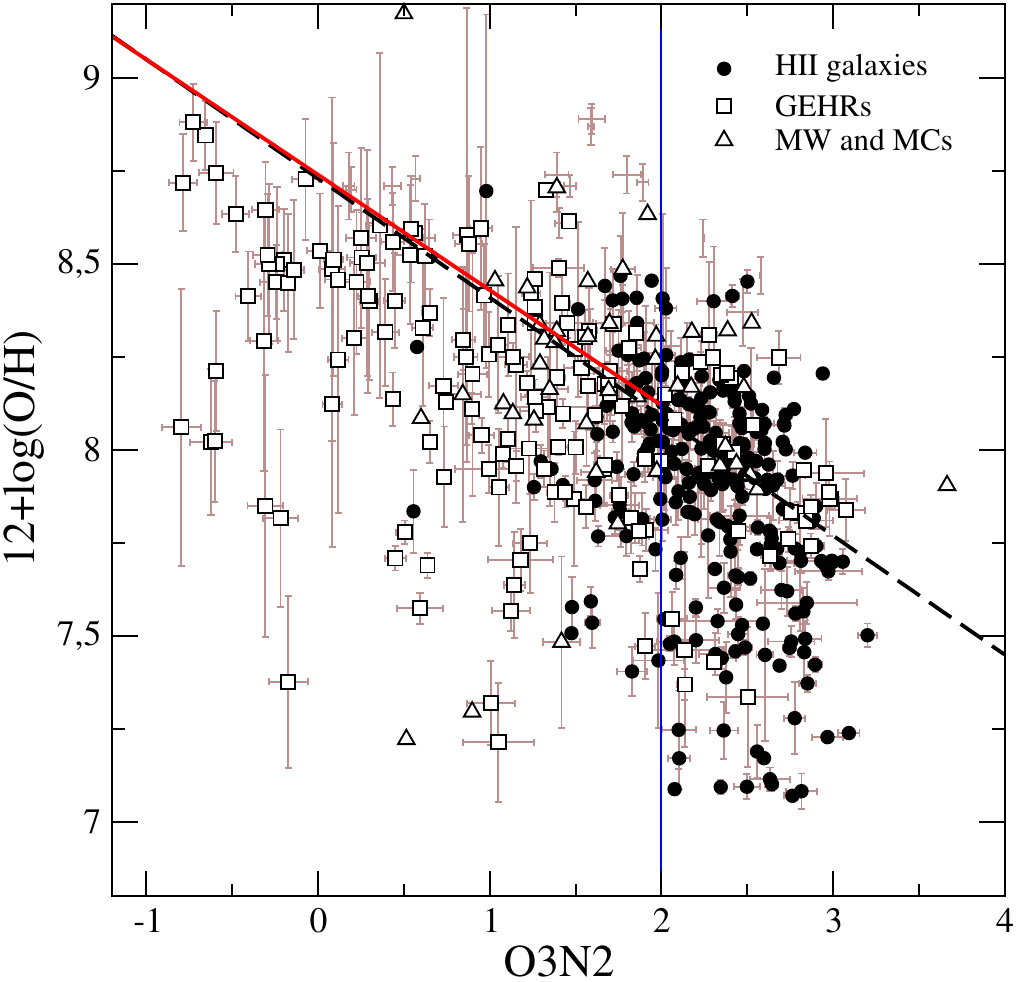}
\hspace*{-1.0cm}
\includegraphics[width=7cm]{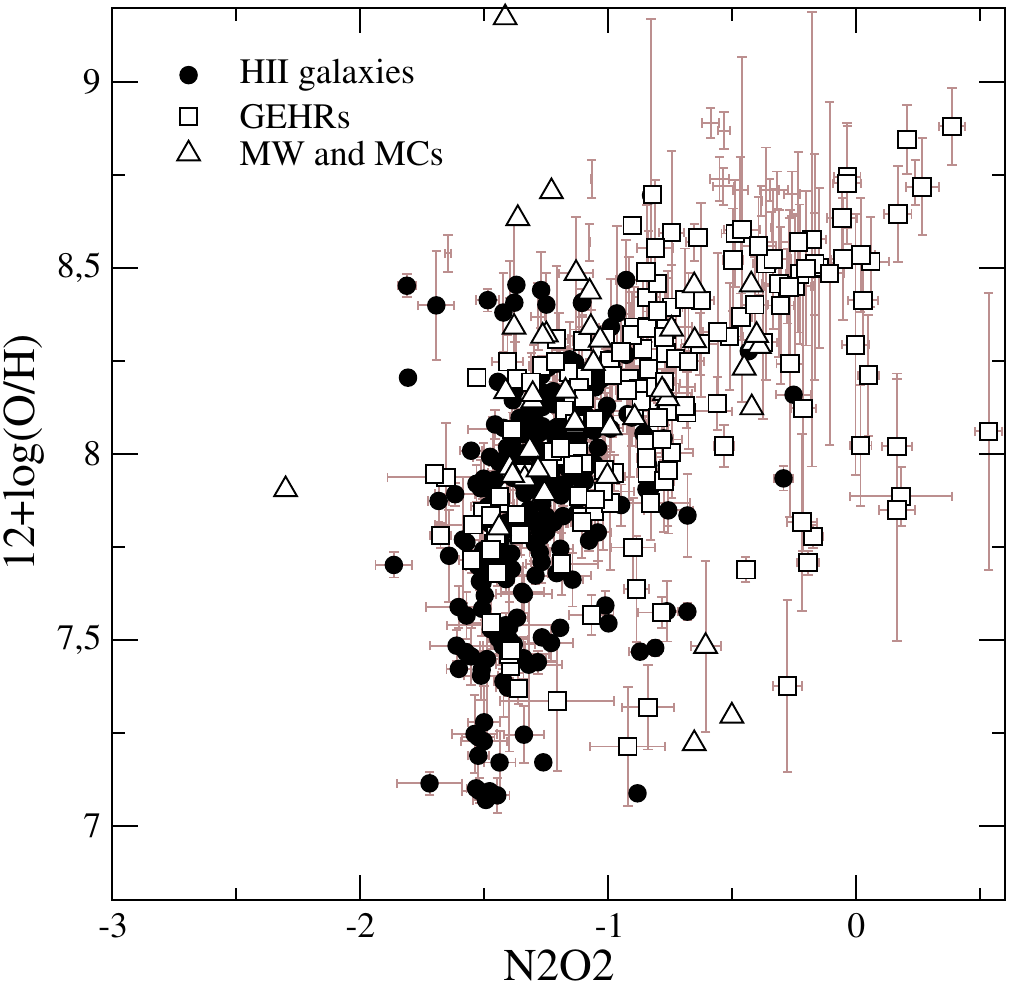}
\includegraphics[width=7cm]{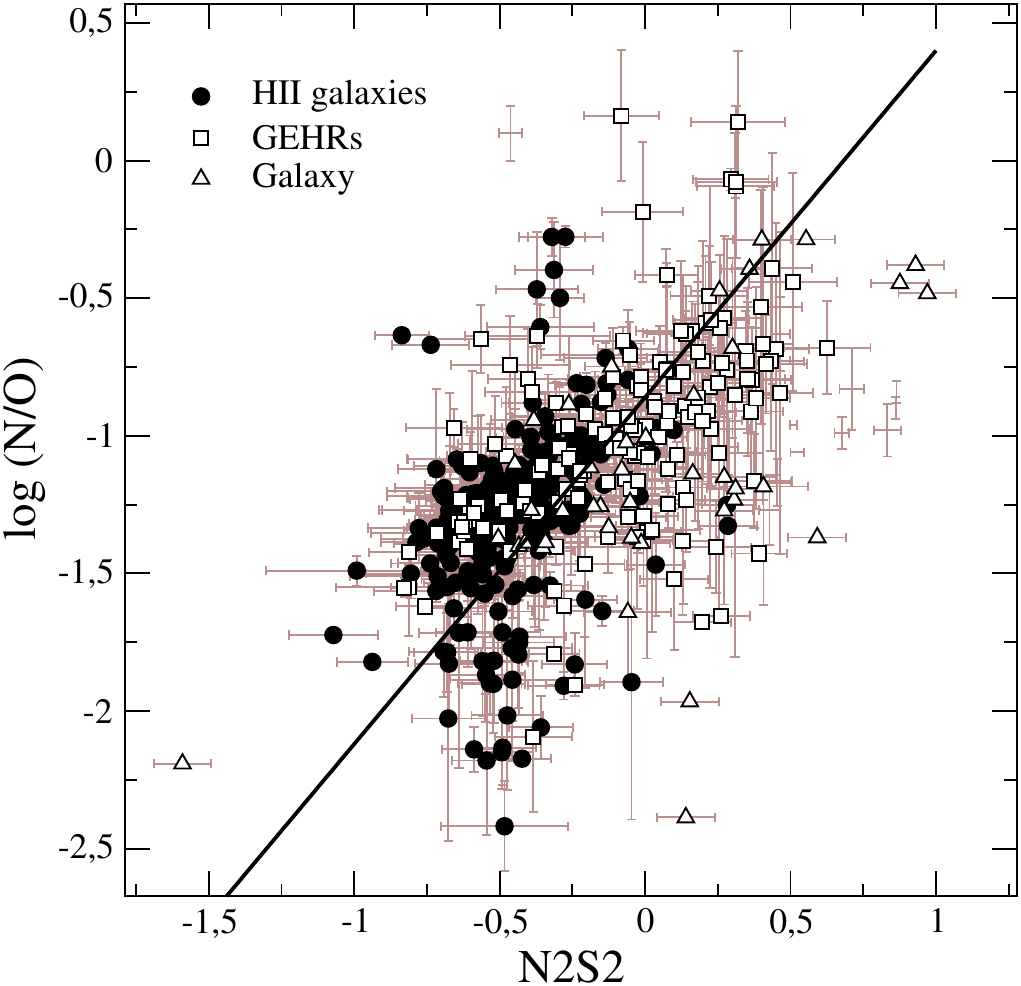}

\caption{Relations between strong-line parameters based on \nii\ emission-lines
and ionic ratios from \cite{pmc09}. From left to right and from up to down:
N2 with O/H, O3N2 with O/H, N2O2 with N/O, and N2S2 with N/O.
}

\label{strong_n}
\end{center}
\end{figure*}

\subsection{Parameters based on \nii\ lines}

The most simple strong line ratio used to derive O/H using \nii\
lines is the N2 parameter, defined by \cite{storchi}:

\begin{equation}
N2 = \log\left(\frac{I(6583)}{I(H\alpha)}\right)
\end{equation}
\\

Notice that, contrary to R23, the log is in the definition of the parameter.
This parameter has at the advantage to be totally independent on
reddening correction or flux calibration uncertainties.
The linear fitting proposed by \cite{pmc09} gives:

\begin{equation}
12+\log(O/H) = 9.07 + 0.79 \cdot N2
\end{equation}
\\

\noindent valid for all the range of metallicity with a dispersion of 0.34 dex.
This relation can be seen in Fig. \ref{strong_n}.
This is usually used to decide to the branch of the R23 parameter, but
the dispersion is very high owing to a large dependence on ionization 
parameter and also to the dispersion in the O/H-N/O relation.

The dependence on log $U$ can be reduced through the definition
of the O3N2 parameter (\cite{alloin79}):

\begin{equation}
O3N2 = \log \left( \frac{I(5007)}{I(H\beta)} \cdot \frac{I(H\alpha)}{I(6583)} \right)
\end{equation}
\\

The linear fitting proposed by \cite{pmc09} leads to:

\begin{equation}
12+\log(O/H) = 8.74 - 0.31 \cdot O3N2
\end{equation}
\\

\noindent with a dispersion of 0.32 dex, but fort
O3N2 $>$ 2.0 as the O3N2 parameter is constant for
lower metallicities, as can be seen in Fig. \ref{strong_n}.

Among other strong emission line ratios the parameter N2O2:

\begin{equation}
N2O2 = \log \left( \frac{I(6583)}{I(3727)}\right)
\end{equation}
\\

\noindent can be used to derive N/O, as can be seen in Fig. \ref{strong_n}.
The linear fitting proposed by \cite{pmc09} gives:

\begin{equation}
\log(N/O) = 0.93 \cdot N2O2 - 0.20
\end{equation}
\\

\noindent with a dispersion of 0.24 dex for all N/O values. A similar parameter
for a lower wavelength baseline is the N2S2 parameter:

\begin{equation}
N2S2 = \log \left( \frac{I(6583)}{I(6717,6731)}\right)
\end{equation}
\\

\noindent whose linear fitting gives:

\begin{equation}
\log(N/O) = 1.26 \cdot N2S2 - 0.86
\end{equation}
\\
	
\noindent with a dispersion of 0.31 dex according to \cite{pmc09}.

\begin{figure*}
\begin{center}
\hspace*{-1.0cm}
\includegraphics[width=7cm]{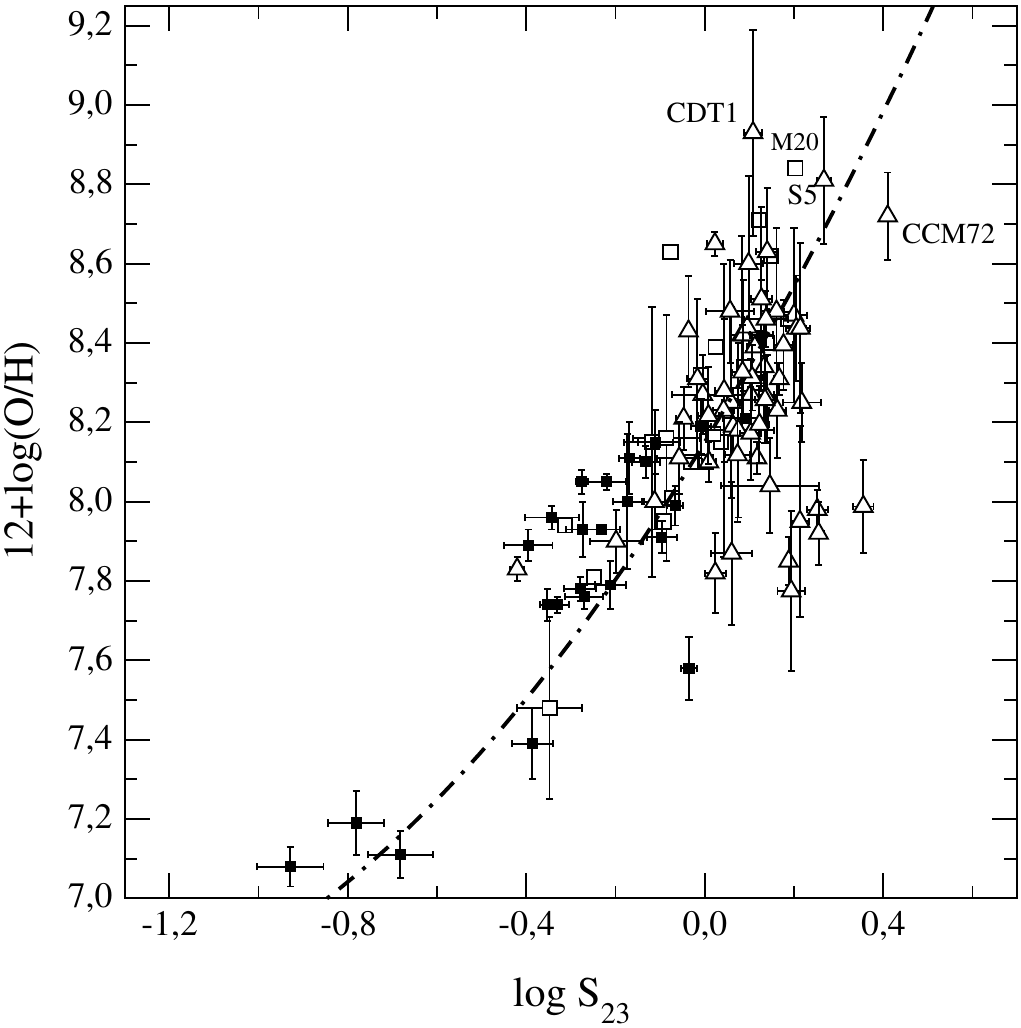}
\includegraphics[width=7cm]{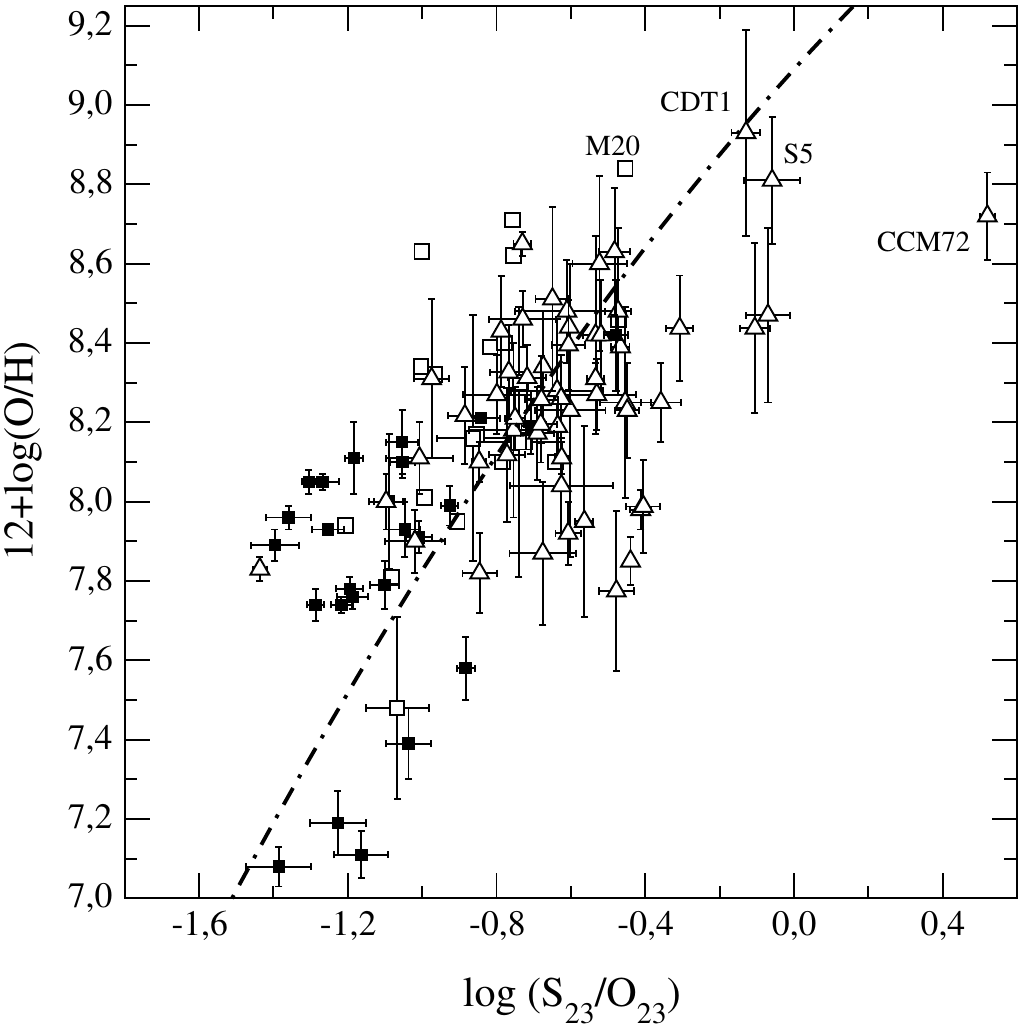}

\caption{Relations and calibrations between the total oxygen abundance and the
S23 parameter (left) and S23/R23 parameter (right) from \cite{pmd05}.
}

\label{strong_s}
\end{center}
\end{figure*}

\subsection{Parameters based on \sii\ and \siii\ lines}

Strong-line methods based on sulfur emission lines provide accurate
oxygen abundances when the red and near-IR spectral optical range is observed and
measured. This is based on the fact that both O and S are primary elements and hence
the S/O ratio is expected to not vary. Following an analogous definition as
for oxygen lines, \cite{ve96} defined the S23 parameter based
on \sii\ and \siii\ lines as:

\begin{equation}
S23 = \frac{I(6717,6731)+I(9099,9532)}{I(H\beta)}
\end{equation}
\\

The polynomial fitting proposed by \cite{pmd05} to derive total
oxygen abundance is:

\begin{equation}
12+\log(O/H) = 8.15 + 0.85\cdot x + 0.58 \cdot x^2
\end{equation}
\\

\noindent where $x$ = log(S23). This can be used up to solar metallicity
(i.e. 12+log(O/H) = 8.69) with a dispersion of 0.20 dex
as the standard deviation of the residuals to the O/H derived from the direct
method.
This relation can be seen in left panel of Fig. \ref{strong_s}.
Similarly \cite{pm06} propose a fitting to derive total sulfur abundances:

\begin{equation}
12+\log(S/H) = 6.622 + 1.860\cdot x + 0.382 \cdot x^2
\end{equation}
\\

\noindent with a dispersion of 0.185 dex in the range of 
-1.0 $\le$ log(S23) $\le$ 0.5.

For high metallicities \cite{pmd05} propose a polynomial fitting to a combination
of both S12 and R23, as can be seen in right panel of Fig. \ref{strong_s}. This fitting gives:

\begin{equation}
12+\log(O/H) = 9.09 + 1.03 \cdot x - 0.22 \cdot x^2
\end{equation}
\\

\noindent where $x$ = log(S23/R23) and the fitting to the objects has a
dispersion of 0.27 dex.

\section{Summary}

In this tutorial it has been reviewed the procedure to analyze the optical emission-line
spectrum from a gaseous nebula ionized by massive star formation. Both recombination
lines emitted from H and He, and collisionally excited emission lines from metallic ions can
be used to derive the physical properties and the ionic and total abundances using the
so-called direct method. 
To do so, it has been provided expressions derived from the software
{\sc pyneb} (\cite{pyneb}) under typical conditions observed in
\hii\ regions and using the most updated sets of atomic coefficients.

The direct method relies on the determination of the electronic 
temperature using the emission-line ratios between strong nebular 
lines and faint auroral emission lines. 
It is also assumed that 
the complete ionization structure is comprised in the observed spectrum 
In this situation one can calculate the
thermal and density radial structure of the distribution  of ionized gas,
measuring as many temperatures as possible and solving
their dependence on density.
Alternatively, one can assume relations between zone temperatures 
from empirical or model-based relations. Then the ionic abundances 
for the corresponding emission lines observed in the spectrum can be 
derived. Finally, the total chemical abundances are calculated by means 
of ionization correction factors that are used to derive the abundances 
of the ions not seen in the observed spectrum

In case that no auroral lines are observed in the optical spectrum, one can resort to
strong-line methods based on the measurement of the nebular lines.
These methods lead formally to determinations of the abundances much less accurate 
than the direct method due to additional dependences of the involved
lines on other functional parameters (e.g. \cite{pmd05}).
However, the most important is that any 
comparison between abundances derived from the direct method and from
strong-line calibrations is done consistently. This can happen
for datasets where the auroral lines are not observed in all points.
In this case strong-line methods calibrated empirically or using models
consistent with the direct method are convenient.

\section*{Acknowledgements}
This tutorial has been done thanks to the financial supports from MINECO Project
AYA2013-47742-C4-1-P and AYA2016-79724-C4-4-P of the Spanish Plan for Astronomy and astrophysics.
I also thank Rub\'en Garc\'\i a-Benito for his kind revision of the manuscript and to
a anonymous referee whose comments have helped to improve the final manuscript.

\end{document}